\documentclass[aps,twocolumn,floatfix,groupedaddress,nofootinbib]{revtex4}
\usepackage{csquotes}
\usepackage{graphicx,color}
\usepackage{float}
\usepackage[caption=false]{subfig}
\usepackage{amsmath}
\usepackage{amssymb}
\usepackage{multirow}
\usepackage{dsfont}
\usepackage{tikz}
\expandafter\let\csname equation*\endcsname\relax
\expandafter\let\csname endequation*\endcsname\relax

\begin{document}

\title{Revisiting the Vashishta--Singwi dielectric scheme for the warm dense uniform electron fluid}
\author{Panagiotis Tolias$^{1}$, Federico Lucco Castello$^{1}$, Fotios Kalkavouras$^{1}$ and Tobias Dornheim$^{2,3}$}
\affiliation{$^1$Space and Plasma Physics - Royal Institute of Technology (KTH), SE-10044 Stockholm, Sweden\\
             $^2$Center for Advanced Systems Understanding (CASUS), D-02826 G\"orlitz, Germany\\
             $^3$Helmholtz-Zentrum Dresden-Rossendorf (HZDR), D-01328 Dresden, Germany}
\begin{abstract}
\noindent The finite temperature version of the Vashishta--Singwi (VS) dielectric scheme for the paramagnetic warm dense uniform electron fluid is revisited correcting for an earlier thermodynamic derivative error. The VS scheme handles quantum mechanical effects at the level of the random phase approximation and treats correlations via the density expansion of a generalized Singwi-Tosi-Land-Sj\"olander (STLS) closure that inserts a parameter determined by enforcing the compressibility sum rule. Systematic comparison with quasi-exact results, based on quantum Monte Carlo simulations, reveals a structural superiority of the VS scheme towards strong coupling and a thermodynamic superiority of the STLS scheme courtesy of a favorable cancellation of errors. Guidelines are provided for the construction of dielectric schemes that are expected to be more accurate but computationally costly.
\end{abstract}
\maketitle

\section{Introduction}

\noindent Warm dense matter (WDM) constitutes an extreme yet ubiquitous state that is characterized by high temperatures ($10^4-10^8\,$K), high pressures ($1-10^4\,$GBar) and beyond-solid densities ($10^{22}-10^{27}\,$cm$^{-3}$)\,\cite{WDMint1,WDMint2,WDMbok1,WDMbok2}. WDM is naturally encountered in dense astrophysical objects (gas giant interiors, dwarf stars, outer neutron star crusts)\,\cite{WDMast1,WDMast2,WDMast3,WDMast4,WDMast5} and is routinely produced in the laboratory (via laser compression, ion beam heating, Z-pinches)\,\cite{WDMlab1,WDMlab2,WDMlab3,WDMlab4,WDMlab5}. Furthermore, the WDM regime is traversed in the early stages of inertial confinement fusion\,\cite{WDMapp1,WDMapp2,WDMapp3} and is also relevant for various technological applications\,\cite{WDMapp4,WDMapp5,WDMapp6,WDMapp7,WDMapp8}. In density-temperature phase diagrams, as the temperature increases, isochoric lines traverse the condense matter region, afterwards the WDM regime and finally the plasma region\,\cite{WDMbok1,WDMbok2,WDMlab1}. This transitional phase diagram location is reflected by the lack of small parameters\,\cite{BoniRev}; the importance of correlations does not allow the use of plasma kinetic theory, the presence of thermal excitations prohibits the straightforward application of solid state approaches and the relevance of quantum effects impedes the use of established theories of classical liquids. Thus, the rigorous theoretical treatment of WDM requires the entire arsenal of quantum many body theory\,\cite{GrossBo,MahanBo,AbrikBo,TDDFTBo,NEqGFBo,CeperBo}; a synthesis that has proven to be notoriously difficult.

The main pre-requisite for the understanding of WDM is the accurate description of the properties of the warm dense uniform electron fluid (UEF), a fundamental homogeneous model system that does not require explicit treatment of the ionic component\,\cite{quanele,DornRev,DornPoP}. For instance, the parametrization of the UEF exchange correlation free energy constitutes essential input for thermal density functional theory (DFT)\,\cite{DFTref1,DFTref2,DFTref3,DFTref4,DFTref5}, while the dynamic local field correction of the UEF is formally equivalent to the exchange correlation kernel of linear-response time-dependent DFT\,\cite{TDDFTr1,TDDFTr2,TDDFTr3} and can be utilized to include correlations into quantum hydrodynamic models\,\cite{QHDref1,QHDref2}.

This has provided the impetus for a worldwide intense research activity targeted at the warm dense UEF\,\cite{DornRev,DornPoP,ESApap6,KaraPRB}. On the one hand, there have been numerous breakthroughs in quantum Monte Carlo (QMC) simulations concerning the invention of entirely new techniques or of novel variants that alleviate the fermion sign problem at different WDM regions\,\cite{newQMC1,newQMC2,newQMC3,newQMC4,newQMC5,newQMC6,newQMC7,newQMC8,newQMC9}. In combination with progress regarding the correction of finite-size errors\,\cite{newQMCa,newQMCb,newQMCc} and the numerical implementation or even full circumvention of analytic continuation\,\cite{newQMCd,newQMCe,newQMCf}, this led to a very accurate description of the thermodynamic, static, dynamic and nonlinear behavior of the UEF in WDM conditions\,\cite{EOSdis1,EOSdis2,ESApap1,ESApap2,WDMsta1,WDMsta2,WDMsta3,WDMsta4}. On the other hand, there have been few developments in finite temperature schemes of the dielectric formalism\,\cite{TanIchi,STLSsch,VSsche1,VSsche2,qSTLSge,qSTLSgr,qSTLSFT,UtsumiI,CASTLSf,HNCSTLS,HNCPIMC,IETChem,IETLett,qIETLet,specia1,specia2,specia3,specia4}; a sophisticated and versatile approach\,\cite{NoziPin,SinTosi,IchiRep,IchiRMP} based on linear density response theory\,\cite{quanele}. In spite of the availability of numerous quasi-exact simulation results and associated parametrizations that can be used for benchmarking, there currently exists no microscopic theoretical approach that can reliably predict the properties of the UEF at WDM conditions.

With this investigation, we aim to improve the theoretical landscape by revisiting the finite temperature version of the Vashishta--Singwi dielectric scheme\,\cite{VSsche1} which supplements the archetypal Singwi-Tosi-Land-Sj\"olander closure\,\cite{STLSsch} with a density expansion that contains an adjustable parameter determined by the exact enforcement of the compressibility sum rule. Our self-consistent formulation corrects for a thermodynamic derivative error that was present in the earlier formulation of Sjostrom and Dufty\,\cite{VSsche2}. An efficient algorithm is devised that is implemented in a hybrid C++/python code. The scheme is numerically solved for thousands of paramagnetic UEF state points that span the WDM regime. The predictions for thermodynamic (interaction energy, exchange correlation free energy) and structural properties (static structure factor, static density response function, pair correlation function, static local field correction) are systematically compared with available quasi-exact results based on path integral Monte Carlo (PIMC) simulations. Finally, we hint at future possibilities for constructing more accurate, albeit more complex, dielectric schemes.

\section{Theoretical}

\noindent The UEF is a spatially homogeneous quantum model system consisting of electrons immersed in a rigid uniform ionic background that is solely characterized by the neutralizing charge density $-ne$, where $n$ is the electron density\,\cite{quanele,DornRev,DornPoP}. It constitutes the quantum mechanical analogue of the classical one-component plasma (OCP)\,\cite{cOCPrev,SIchRMP,OCPbrid}, thus it is often also referred to as the quantum OCP. The UEF thermodynamic state points are specified by three dimensionless parameters\,\cite{quanele,DornRev,DornPoP}: \textbf{(1)} the \emph{quantum coupling parameter} $r_{\mathrm{s}}=d/a_{\mathrm{B}}$ with $d$ the Wigner-Seitz radius $d=(4\pi{n}/3)^{-1/3}$ and $a_{\mathrm{B}}=\hbar^2/(m_{\mathrm{e}}e^2)$ the Bohr radius, \textbf{(2)} the \emph{quantum degeneracy parameter} $\Theta=T/E_{\mathrm{F}}$ with $E_{\mathrm{F}}=[(6\pi^2n^{\uparrow})^{2/3}/2](\hbar^2/m_{\mathrm{e}})$ the Fermi energy with respect to the Fermi wave-vector of the spin-up electrons $k_{\mathrm{F}}^{\uparrow}=(6\pi^2n^{\uparrow})^{1/3}$ and $T$ the temperature in energy units, \textbf{(3)} the \emph{spin polarization parameter} $\xi=(n^{\uparrow}-n^{\downarrow})/n$ for which $0\leq\xi\leq1$ within the standard convention $n^{\uparrow}\geq{n}^{\downarrow}$. In this work, we exclusively study the paramagnetic (or unpolarized) case of equal spin-up and -down electrons, $\xi=0$. It is noted that, in contrast to the UEF, OCP state points are specified by a unique dimensionless parameter\,\cite{cOCPrev,SIchRMP,OCPbrid}: the classical coupling parameter $\Gamma=e^2/(dT)$ for which $\Gamma=2\lambda^2r_{\mathrm{s}}/\Theta$ with $\lambda^3=(k_{\mathrm{F}}d)^{-3}=4/(9\pi)$.

Within the UEF phase diagram, the WDM regime can be roughly demarcated by $0.1\lesssim{r}_{\mathrm{s}},\Theta\lesssim10$\,\cite{DornRev,DornPoP}. The lack of small parameters, $r_{\mathrm{s}}\sim\Theta\sim1$, is indicative of the complex interplay between quantum effects (exchange, diffraction), bare Coulomb interactions and thermal excitations which make the theoretical treatment of the warm dense UEF a formidable task.

In the high density degenerate limit $r_{\mathrm{s}}\to0$, the UEF approaches the non-interacting Fermi gas\,\cite{quanele}. The ideal (or Lindhard) density response of the free unpolarized electrons is given by\,\cite{quanele,DornRev}
\begin{equation}
\chi_0(\boldsymbol{k},\omega)=2\int\frac{d^3q}{(2\pi)^3}\frac{f_0\left(\boldsymbol{q}\right)-f_0\left(\boldsymbol{q}+\boldsymbol{k}\right)}{\hbar\omega+\epsilon(\boldsymbol{q})-\epsilon(\boldsymbol{q}+\boldsymbol{k})+\imath0}\,,\label{Lindhard}
\end{equation}
with $\epsilon(\boldsymbol{q})=\hbar^2q^2/(2m_{\mathrm{e}})$ the electron kinetic energy and $f_0(\boldsymbol{q})$ the Fermi-Dirac distribution function
\begin{equation}
{f}_0\left(\boldsymbol{q}\right)=\frac{1}{\exp{\left(\displaystyle\frac{\hbar^2q^2}{2m_{\mathrm{e}}T}-\bar{\mu}\right)}+1}\,,\label{FermiDirac}
\end{equation}
with $\bar{\mu}=\beta\mu$ the reduced chemical potential ($\mu$ is the chemical potential and $1/\beta=T$) that is determined by the normalization condition $\int[d^3q/(2\pi)^3]{f}_0\left(\boldsymbol{q}\right)=n/2$.

In the low density classical limit $\Gamma\to0$, the UEF approaches the non-interacting classical gas\,\cite{quanele}. The ideal (or Vlasov) density response of the free electrons is then given by\,\cite{IchimaI,OCPdiel}
\begin{equation}
\chi_0(\boldsymbol{k},\omega)=-\int\,d^3p\frac{1}{\omega-\boldsymbol{k}\cdot\boldsymbol{v}+\imath0}\left[\boldsymbol{k}\cdot\frac{\partial{f}_0(\boldsymbol{p})}{\partial\boldsymbol{p}}\right]\,,\label{Vlasov}
\end{equation}
where ${f}_0(\boldsymbol{p})$ is the Maxwellian distribution in momentum space that is normalized according to $\int\,d^3p{f}_0\left(\boldsymbol{p}\right)=n$.

\subsection{Dielectric formalism}\label{subsec:gendiel}

\noindent The self-consistent dielectric formalism constitutes one of the most accurate and versatile microscopic frameworks for the description of the thermodynamic and static properties of interacting uniform quantum systems such as the UEF\,\cite{NoziPin,SinTosi,IchiRep,quanele,IchiRMP,DornRev,DornPoP}. It combines fundamental results of the density version of linear response theory\,\cite{quanele} with an approximate closure stemming from perturbative quantum/classical kinetic theories of non-ideal gases\,\cite{quankin,IchimaB} or from non-perturbative integral equation theories / memory function approaches of classical liquids\,\cite{IchimaB,IETliqu}.

In the polarization potential approach\,\cite{IchimaB}, the density response function $\chi(\boldsymbol{k},\omega)$ is expressed very generally in terms of the ideal (Lindhard) density response $\chi_0(\boldsymbol{k},\omega)$ and the dynamic local field correction $G(\boldsymbol{k},\omega)$ (LFC) as
\begin{equation}
\chi(\boldsymbol{k},\omega)=\frac{\chi_0(\boldsymbol{k},\omega)}{1-U(\boldsymbol{k})\left[1-G(\boldsymbol{k},\omega)\right]\chi_0(\boldsymbol{k},\omega)}\,,\label{densityresponseDLFC}
\end{equation}
where $U(\boldsymbol{k})$ is the Fourier transformed pair interaction energy, with $U(\boldsymbol{k})=4\pi{e}^2/k^2$ for Coulomb interactions. In addition, for finite temperature systems, the combination of the zero frequency moment sum rule, the quantum fluctuation--dissipation theorem and the analytic continuation of the causal $\chi(\boldsymbol{k},\omega)$ to the complex frequency plane yield a static structure factor $S(\boldsymbol{k})$ (SSF) relation that involves the infinite Matsubara summation\,\cite{TanIchi}
\begin{equation}
S(\boldsymbol{k})=-\frac{1}{{n}\beta}\displaystyle\sum_{l=-\infty}^{\infty}\widetilde{\chi}(\boldsymbol{k},\imath\omega_l)\,,\label{Matsubaraseries}
\end{equation}
with $\widetilde{\chi}(\boldsymbol{k},z)$ the analytically-continued density response function and $\omega_l=2\pi{l}/(\beta\hbar)$ the bosonic Matsubara frequencies. Finally, the dynamic LFC, which incorporates Pauli exchange, quantum diffraction and Coulomb correlation effects beyond the mean field description, is given by a complicated SSF functional of the general form
\begin{equation}
G(\boldsymbol{k},\omega)\equiv{G}[S](\boldsymbol{k},\omega)\,.\label{functionalclosure}
\end{equation}
The combination of Eqs.(\ref{densityresponseDLFC},\ref{Matsubaraseries},\ref{functionalclosure}) leads to a non-linear functional equation of the type
\begin{equation}
S(\boldsymbol{k})=-\frac{1}{{n}\beta}\displaystyle\sum_{l=-\infty}^{\infty}\frac{\widetilde{\chi}_0(\boldsymbol{k},\imath\omega_l)}{1-U(\boldsymbol{k})\left[1-{G}[S](\boldsymbol{k},\imath\omega_l)\right]\widetilde{\chi}_0(\boldsymbol{k},\imath\omega_l)}\,,\nonumber
\end{equation}
to be solved for the SSF\,\cite{DornRev,TanIchi}.

Multiple dielectric schemes have been developed over the last six decades that differ in the approximate treatment of the exact (unknown) LFC functional. They can be broadly categorized in the following manner; \textbf{(i)} semi-classical schemes based on classical kinetic equations\,\cite{STLSsch,VSsche1,VSsche2,TanIchi}, \textbf{(ii)} pure quantum schemes based on quantum kinetic equations\,\cite{qSTLSge,qSTLSgr,qSTLSFT,UtsumiI}, \textbf{(iii)} semi-classical schemes based on various integral equation theories\,\cite{CASTLSf,HNCSTLS,HNCPIMC,IETChem,IETLett}, \textbf{(iv)} quantized schemes also based on integral equation theories\,\cite{qIETLet}, \textbf{(v)} memory-function or viscoelastic schemes incorporating the elusive third frequency moment sum rule\,\cite{specia1,specia2,specia3,specia4}, \textbf{(vi)} semi-empirical schemes based on exact QMC results and incorporating known asymptotic limits\,\cite{ESApap1,ESApap2}.

\subsection{Vashishta--Singwi scheme: General formulation}\label{subsec:genVS}

\noindent Vashishta--Singwi (VS) theory is a semi-classical scheme of the self-consistent dielectric formalism\,\cite{VSsche1}. The treatment of correlations is classical, since the $s=1$ member of the classical Bogoliubov-Born-Green-Kirkwood-Yvon (BBGKY) hierarchy of s-reduced distribution functions is truncated with an equilibrium factorization Ansatz. The treatment of quantum mechanical effects is restricted to the random phase approximation level, since the resulting ideal Vlasov density response is substituted with the ideal Lindhard density response. It is noted that equilibrium closures to the classical BBGKY hierarchy imply a static LFC (SLFC), thus the VS closure leads to a SLFC.

The VS closure of the classical BBGKY hierarchy reads as $f_2(\boldsymbol{r},\boldsymbol{p},\boldsymbol{r}^{\prime},\boldsymbol{p}^{\prime},t)=f(\boldsymbol{r},\boldsymbol{p},t)f(\boldsymbol{r}^{\prime},\boldsymbol{p}^{\prime},t)g(\boldsymbol{r},\boldsymbol{r}^{\prime},t)$ where $f_2$ is the two-particle distribution function, $f$ is the one-particle distribution function and $g(\boldsymbol{r},\boldsymbol{r}^{\prime},t)=g_{\mathrm{eq}}(|\boldsymbol{r}-\boldsymbol{r}^{\prime}|;n,T)+{\alpha}\left[\delta{n}(\boldsymbol{r},t)+\delta{n}(\boldsymbol{r}^{\prime},t)\right]\left[\partial{g}_{\mathrm{eq}}(|\boldsymbol{r}-\boldsymbol{r}^{\prime}|;n,T)/\partial{n}\right]$ is a non-equilibrium pair correlation function (PCF) with $g_{\mathrm{eq}}$ the equilibrium PCF within the assumptions of a homogeneous and isotropic system. The VS closure stems from the long-time long-range classical limit $g(\boldsymbol{r},\boldsymbol{r}^{\prime},t)=g_{\mathrm{eq}}(|\boldsymbol{r}-\boldsymbol{r}^{\prime}|;n,T)+\delta{n}\left[\partial{g}_{\mathrm{eq}}(|\boldsymbol{r}-\boldsymbol{r}^{\prime}|;n,T)/\partial{n}\right]$ in absence of temperature perturbations, which is first extended to arbitrary spatial scales via $g(\boldsymbol{r},\boldsymbol{r}^{\prime},t)=g_{\mathrm{eq}}(|\boldsymbol{r}-\boldsymbol{r}^{\prime}|;n,T)+(1/2)\left[\delta{n}(\boldsymbol{r},t)+\delta{n}(\boldsymbol{r}^{\prime},t)\right]\left[\partial{g}_{\mathrm{eq}}(|\boldsymbol{r}-\boldsymbol{r}^{\prime}|;n,T)/\partial{n}\right]$ and is then generalized to the quantum case by the ad-hoc insertion of the parameter $\alpha\equiv\alpha(n,T)$\,\cite{VSsche1,VSschep}. This parameter is state point-dependent and is often referred to as the self-consistency parameter, since it is determined by requiring that the compressibility sum rule (CSR)\,\cite{quanele,DornRev,DornPoP,IchiRMP} is satisfied exactly. It is emphasized that the choice of $\alpha=0$ directly leads to the archetypal Singwi-Tosi-Land-Sj\"olander (STLS) closure that reads as $f_2(\boldsymbol{r},\boldsymbol{p},\boldsymbol{r}^{\prime},\boldsymbol{p}^{\prime},t)=f(\boldsymbol{r},\boldsymbol{p},t)f(\boldsymbol{r}^{\prime},\boldsymbol{p}^{\prime},t)g_{\mathrm{eq}}(|\boldsymbol{r}-\boldsymbol{r}^{\prime}|;n,T)$\,\cite{STLSsch}.

After substitution of the VS closure to the first member of the classical BBGKY hierarchy and use of $g_{\mathrm{eq}}(|\boldsymbol{r}-\boldsymbol{r}^{\prime}|)=1+h(|\boldsymbol{r}-\boldsymbol{r}^{\prime}|)$ with $h(\cdot)$ the total correlation function, the VS kinetic equation reads as
\begin{widetext}
\begin{align*}
&\left\{\frac{\partial}{\partial{t}}+\boldsymbol{v}\cdot\frac{\partial}{\partial\boldsymbol{r}}-\frac{\partial}{\partial\boldsymbol{r}}U_{\mathrm{ext}}(\boldsymbol{r},t)\cdot\frac{\partial}{\partial\boldsymbol{p}}\right\}f(\boldsymbol{r},\boldsymbol{p},t)=\frac{\partial}{\partial\boldsymbol{r}}\left\{\int{U}(\boldsymbol{r}-\boldsymbol{r}^{\prime})n(\boldsymbol{r}^{\prime},t)d^3r^{\prime}\right\}\cdot\frac{\partial{f}(\boldsymbol{r},\boldsymbol{p},t)}{\partial\boldsymbol{p}}+\nonumber\\&\frac{\partial}{\partial\boldsymbol{r}}\left\{\int{U}(\boldsymbol{r}-\boldsymbol{r}^{\prime})h(|\boldsymbol{r}-\boldsymbol{r}^{\prime}|)n(\boldsymbol{r}^{\prime},t)d^3r^{\prime}\right\}\cdot\frac{\partial{f}(\boldsymbol{r},\boldsymbol{p},t)}{\partial\boldsymbol{p}}+\frac{\partial}{\partial\boldsymbol{r}}\left\{{\alpha}\int{U}(\boldsymbol{r}-\boldsymbol{r}^{\prime})\delta{n}(\boldsymbol{r}^{\prime},t)\frac{\partial{h}(|\boldsymbol{r}-\boldsymbol{r}^{\prime}|)}{\partial{n}}n(\boldsymbol{r}^{\prime},t)d^3r^{\prime}\right\}\cdot\frac{\partial{f}(\boldsymbol{r},\boldsymbol{p},t)}{\partial\boldsymbol{p}}\nonumber\\&+\frac{\partial}{\partial\boldsymbol{r}}\left\{{\alpha}\int{U}(\boldsymbol{r}-\boldsymbol{r}^{\prime})\delta{n}(\boldsymbol{r},t)\frac{\partial{h}(|\boldsymbol{r}-\boldsymbol{r}^{\prime}|)}{\partial{n}}n(\boldsymbol{r}^{\prime},t)d^3r^{\prime}\right\}\cdot\frac{\partial{f}(\boldsymbol{r},\boldsymbol{p},t)}{\partial\boldsymbol{p}}\,,\nonumber
\end{align*}
\end{widetext}
with $U_{\mathrm{ext}}(\boldsymbol{r},t)$ the external potential energy and $U(\boldsymbol{r}-\boldsymbol{r}^{\prime})$ the pair interaction potential energy. At the RHS, the first bracketed term represents the mean field contribution, the second bracketed term represents the STLS polarization field contribution and the last two bracketed terms correspond to the VS polarization field corrections. Application of an external potential energy perturbation to the system (initially in equilibrium), linearization with respect to the small perturbation strength, use of spatio-temporal Fourier transforms, consideration of the adiabatic switching of the perturbation and substitution for the SSF via $S(\boldsymbol{k})=1+nH(\boldsymbol{k})$ lead to
\begin{widetext}
\begin{align*}
&\delta{f}^{\boldsymbol{p}}_{\boldsymbol{k},\omega}=\left[-\frac{\boldsymbol{k}}{\omega-\boldsymbol{k}\cdot\boldsymbol{v}+\imath0}\cdot\frac{\partial{f}_0(\boldsymbol{p})}{\partial\boldsymbol{p}}\right]\left\{\delta{U}^{\mathrm{ext}}_{\boldsymbol{k},\omega}+U(\boldsymbol{k})\left\{1-\left[1+{\alpha}n\frac{\partial}{\partial{n}}\right]\left[-\frac{1}{n}\int\frac{\boldsymbol{k}\cdot\boldsymbol{k}^{\prime}}{k^2}\frac{{U}(\boldsymbol{k}^{\prime})}{{U}(\boldsymbol{k})}\left[S(|\boldsymbol{k}-\boldsymbol{k}^{\prime}|)-1\right]\frac{d^3k^{\prime}}{(2\pi)^3}\right]\delta{n}_{\boldsymbol{k},\omega}\right\}\right\}\,.\nonumber
\end{align*}
\end{widetext}
Integration over the momenta to obtain the density perturbation $\delta{n}_{\boldsymbol{k},\omega}=\int\delta{f}^{\boldsymbol{p}}_{\boldsymbol{k},\omega}d^3p$, solution of the resulting linear equation with respect to the density perturbation $\delta{n}_{\boldsymbol{k},\omega}$,  identification of the ideal Vlasov density response $\chi_0(\boldsymbol{k},\omega)=-\int\boldsymbol{k}\cdot[\partial{f}_0(\boldsymbol{p})/\partial\boldsymbol{p}]/[\omega-\boldsymbol{k}\cdot\boldsymbol{v}+\imath0]d^3p$ and use of the functional derivative definition of the linear density response function $\chi(\boldsymbol{k},\omega)=\delta{n}_{\boldsymbol{k},\omega}/\delta{U}_{\boldsymbol{k},\omega}^{\mathrm{ext}}$ yield the VS density response function
\begin{widetext}
\begin{align*}
\chi(\boldsymbol{k},\omega)=\displaystyle\frac{\chi_0(\boldsymbol{k},\omega)}{1-U(\boldsymbol{k})\left\{1-\left[1+{\alpha}n\displaystyle\frac{\partial}{\partial{n}}\right]\left[-\displaystyle\frac{1}{n}\int\displaystyle\frac{\boldsymbol{k}\cdot\boldsymbol{k}^{\prime}}{k^2}\displaystyle\frac{{U}(\boldsymbol{k}^{\prime})}{{U}(\boldsymbol{k})}\left[S(|\boldsymbol{k}-\boldsymbol{k}^{\prime}|)-1\right]\frac{d^3k^{\prime}}{(2\pi)^3}\right]\right\}\chi_0(\boldsymbol{k},\omega)}\,.\nonumber
\end{align*}
\end{widetext}
After term-by-term comparison with the general form of the density response function, Eq.(\ref{densityresponseDLFC}), and substitution for the Fourier transformed Coulomb interaction $U(\boldsymbol{k})=4\pi{e}^2/k^2$, it is evident that the VS SLFC is given by
\begin{align*}
G_{\mathrm{VS}}(\boldsymbol{k})=\left[1+{\alpha}n\displaystyle\frac{\partial}{\partial{n}}\right]\left[-\displaystyle\frac{1}{n}\int\displaystyle\frac{\boldsymbol{k}\cdot\boldsymbol{k}^{\prime}}{{k^{\prime}}^2}\left[S(|\boldsymbol{k}-\boldsymbol{k}^{\prime}|)-1\right]\frac{d^3k^{\prime}}{(2\pi)^3}\right]\,.\nonumber
\end{align*}
The second bracketed factor can be identified to be equal to the STLS SLFC. Thus, the VS SLFC can be compactly rewritten as
\begin{align}
&G_{\mathrm{VS}}(\boldsymbol{k})=\left(1+{\alpha}n\displaystyle\frac{\partial}{\partial{n}}\right)G_{\mathrm{STLS}}(\boldsymbol{k})\,,\label{VSSLFCgen}\\
&G_{\mathrm{STLS}}(\boldsymbol{k})=-\displaystyle\frac{1}{n}\int\displaystyle\frac{\boldsymbol{k}\cdot\boldsymbol{k}^{\prime}}{{k^{\prime}}^2}\left[S(|\boldsymbol{k}-\boldsymbol{k}^{\prime}|)-1\right]\frac{d^3k^{\prime}}{(2\pi)^3}\,.\label{STLSSLFCgen}
\end{align}

Finally, the CSR is most generally expressed through the long-wavelength limit of the static density response (SDR) as\,\cite{quanele,IchiRMP,SinTosi}
\begin{align*}
\displaystyle\lim_{k\to0}\left[\frac{1}{\chi(\boldsymbol{k},0)}+U(\boldsymbol{k})\right]=-\frac{\partial^2}{\partial{n}^2}[nf(n,T)]\,,\nonumber
\end{align*}
where $f(n,T)$ denotes the total free energy per particle. Application of the above formula for the interacting and non-interacting case, subtraction by parts, substitution for the general density response function from Eq.(\ref{densityresponseDLFC}) and introduction of the exchange-correlation free energy (per particle) $f_{\mathrm{xc}}$, allows to alternatively express the CSR via the long-wavelength limit of the SLFC as\,\cite{DornRev,DornPoP}
\begin{align}
\displaystyle\lim_{k\to0}\frac{G_{\mathrm{VS}}(\boldsymbol{k})}{k^2}=-\frac{1}{4\pi{e}^2}\frac{\partial^2}{\partial{n}^2}[nf_{\mathrm{xc}}(n,T)]\,.\label{VSCSRgen}
\end{align}
Taking into account that $G_{\mathrm{VS}}(\boldsymbol{k})$ explicitly depends on $\alpha$ and that $f_{\mathrm{xc}}(n,T)$ implicity depends on $\alpha$, Eq.(\ref{VSCSRgen}) essentially constitutes a non-linear equation that determines the self-consistency parameter $\alpha$ for every state point. It is also important to emphasize that the $1/2\leq\alpha\lesssim1$ UEF ground state bounds\,\cite{VSsche1} and the $\alpha\simeq2/3$ UEF ground state approximation\,\cite{VSsche1} are not applicable to the finite temperature UEF.

\subsection{Vashishta--Singwi scheme: Correct finite temperature formulation}\label{subsec:FTVS}

\noindent In the standard normalized units, for the correct calculation of the density derivative present in the VS SLFC, one should take into account that the STLS SLFC depends on the density through the quantum coupling parameter $r_{\mathrm{s}}$ (see the Wigner-Seitz radius), the quantum degeneracy parameter $\Theta$ (see the Fermi energy) and the normalized wavenumber $x=k/k_{\mathrm{F}}$ (see the Fermi wavenumber). Thus, the chain differentiation rule leads to the VS SLFC
\begin{align}
G_{\mathrm{VS}}(x;r_{\mathrm{s}},\Theta)&=\left[1+\alpha(r_{\mathrm{s}},\Theta)\left(-\frac{2}{3}\Theta\frac{\partial}{\partial\Theta}-\frac{1}{3}r_{\mathrm{s}}\frac{\partial}{\partial{r}_{\mathrm{s}}}\right.\right.\nonumber\\&\quad\left.\left.-\frac{1}{3}x\frac{\partial}{\partial{x}}\right)\right]{G}_{\mathrm{STLS}}(x;r_{\mathrm{s}},\Theta)\,,\label{VSSLFCftcorr}
\end{align}
with the STLS SLFC given by\,\cite{TanIchi,DornRev}
\begin{align}
G_{\mathrm{STLS}}(x;r_{\mathrm{s}},\Theta)&=-\frac{3}{4}\int_0^{\infty}dyy^2\left[S(y;r_{\mathrm{s}},\Theta)-1\right]\nonumber\\&\quad\times\left[1+\frac{x^2-y^2}{2xy}\ln{\left|\frac{x+y}{x-y}\right|}\right]\,.\label{STLSSLFCftcorr}
\end{align}
In the standard normalized units where the energies are normalized with the Hartree energy, the CSR becomes
\begin{align*}
&\lim_{x\to0}\frac{G_{\mathrm{VS}}(x;r_{\mathrm{s}},\Theta)}{x^2}=-\frac{3\pi}{4}\lambda{r}_{\mathrm{s}}\left\{n\frac{\partial^2}{\partial{n}^2}\left[n\widetilde{f}_{\mathrm{xc}}(r_{\mathrm{s}},\Theta)\right]\right\}\,.\nonumber
\end{align*}
The RHS is first computed by taking into account that the exchange-correlation free energy depends on the density through the quantum coupling parameter $r_{\mathrm{s}}$ (see the Wigner-Seitz radius) and the quantum degeneracy parameter $\Theta$ (see the Fermi energy). Thus, repeated applications of the chain differentiation rule yield
\begin{align*}
n\frac{\partial^2}{\partial{n}^2}\left[n\widetilde{f}_{\mathrm{xc}}(r_{\mathrm{s}},\Theta)\right]&=\frac{1}{9}\left(4\Theta^2\frac{\partial^2}{\partial{\Theta}^2}+r_{\mathrm{s}}^2\frac{\partial^2}{\partial{r_{\mathrm{s}}^2}}+4\Theta{r}_{\mathrm{s}}\frac{\partial^2}{\partial{\Theta}\partial{r_{\mathrm{s}}}}\right.\nonumber\\&\quad\left.-2\Theta\frac{\partial}{\partial{\Theta}}-2r_{\mathrm{s}}\frac{\partial}{\partial{r_{\mathrm{s}}}}\right)\widetilde{f}_{\mathrm{xc}}(r_{\mathrm{s}},\Theta)\,.\nonumber
\end{align*}
The LHS is then computed with the aid of the straightforward [see Eq.(\ref{STLSSLFCftcorr})] long-wavelength limits
\begin{align*}
&\lim_{x\to0}\frac{G_{\mathrm{STLS}}(x;r_{\mathrm{s}},\Theta)}{x^2}=-\frac{1}{2}\pi\lambda{r}_{\mathrm{s}}\widetilde{u}_{\mathrm{int}}(r_{\mathrm{s}},\Theta)\,,\nonumber\\
&\lim_{x\to0}\frac{1}{x}\frac{\partial}{\partial{x}}G_{\mathrm{STLS}}(x;r_{\mathrm{s}},\Theta)=-\pi\lambda{r}_{\mathrm{s}}\widetilde{u}_{\mathrm{int}}(r_{\mathrm{s}},\Theta)\,,\nonumber
\end{align*}
where the fundamental expression for the interaction energy of the UEF\,\cite{quanele,IchiRMP,DornRev} has been employed
\begin{align}
\widetilde{u}_{\mathrm{int}}(r_{\mathrm{s}},\Theta)=\frac{1}{\pi\lambda{r}_{\mathrm{s}}}\int_0^{\infty}dy[S(y;r_{\mathrm{s}},\Theta)-1]\,.\label{interaction_energy}
\end{align}
Substituting for the VS SLFC from Eq.(\ref{VSSLFCftcorr}) and utilizing the above two limits, one expresses the LHS as
\begin{align*}
\lim_{x\to0}\frac{G_{\mathrm{VS}}(x;r_{\mathrm{s}},\Theta)}{x^2}&=-\frac{\pi}{2}\lambda{r}_{\mathrm{s}}\left\{1-\alpha(r_{\mathrm{s}},\Theta)\left[1+\frac{2}{3}\Theta\frac{\partial}{\partial\Theta}\right.\right.\nonumber\\&\quad\left.\left.+\frac{1}{3}r_{\mathrm{s}}\frac{\partial}{\partial{r}_{\mathrm{s}}}\right]\right\}\widetilde{u}_{\mathrm{int}}(r_{\mathrm{s}},\Theta)\,.
\end{align*}
Finally, combining the RHS, LHS expressions with the CSR expression, solving the linear explicit dependence with respect to the self-consistency parameter $\alpha(r_{\mathrm{s}},\Theta)$, utilizing the differential version of the adiabatic connection formula in order to symmetrize the numerators and denominators\,\cite{quanele,IchiRMP,DornRev}
\begin{align}
&\widetilde{f}_{\mathrm{xc}}(r_{\mathrm{s}},\Theta)=\frac{1}{r^2_{\mathrm{s}}}\int_0^{r_{\mathrm{s}}}r_{\mathrm{s}}'\widetilde{u}_{\mathrm{int}}(r_{\mathrm{s}}',\Theta)dr_{\mathrm{s}}'\,,\label{adiabaticInt}\\
&\widetilde{u}_{\mathrm{int}}(r_{\mathrm{s}},\Theta)=2\widetilde{f}_{\mathrm{xc}}(r_{\mathrm{s}},\Theta)+r_{\mathrm{s}}\frac{\partial}{\partial{r}_{\mathrm{s}}}\widetilde{f}_{\mathrm{xc}}(r_{\mathrm{s}},\Theta)\,,\label{adiabaticDif}
\end{align}
and carrying out some simple algebraic manipulations, the CSR expression for the finite temperature VS scheme is ultimately transformed to a compact equation for the self-consistency parameter $\alpha$. It is emphasized that this equation is highly non-linear owing to the implicit dependence of the interaction energy and exchange-correlation energy on $\alpha$, as discerned by inspecting Eqs.(\ref{interaction_energy},\ref{adiabaticInt}) in view of the $\alpha-$dependent VS SLFC, see Eqs.(\ref{VSSLFCftcorr},\ref{STLSSLFCftcorr}). The compact equation reads as
\begin{widetext}
\begin{align}
&\alpha(r_{\mathrm{s}},\Theta)=\frac{\displaystyle\left(2-\frac{2}{3}\Theta^2\frac{\partial^2}{\partial{\Theta}^2}-\frac{1}{6}r_{\mathrm{s}}^2\frac{\partial^2}{\partial{r_{\mathrm{s}}^2}}-\frac{2}{3}\Theta{r}_{\mathrm{s}}\frac{\partial^2}{\partial{\Theta}\partial{r_{\mathrm{s}}}}+\frac{1}{3}\Theta\frac{\partial}{\partial{\Theta}}+\frac{4}{3}r_{\mathrm{s}}\frac{\partial}{\partial{r_{\mathrm{s}}}}\right)\widetilde{f}_{\mathrm{xc}}(r_{\mathrm{s}},\Theta)}{\displaystyle\left(1+\frac{2}{3}\Theta\frac{\partial}{\partial\Theta}+\frac{1}{3}r_{\mathrm{s}}\frac{\partial}{\partial{r}_{\mathrm{s}}}\right)\widetilde{u}_{\mathrm{int}}(r_{\mathrm{s}},\Theta)}\,.\label{selfconsistencyVS}
\end{align}
\end{widetext}
To sum up, in normalized units, the correct finite temperature formulation of the VS closure consists of Eqs.(\ref{VSSLFCftcorr},\ref{STLSSLFCftcorr}) for the SLFC and Eq.(\ref{selfconsistencyVS}) for the CSR.

\subsection{Vashishta--Singwi scheme: Sjostrom--Dufty finite temperature formulation}\label{subsec:SDVS}

\noindent As aforementioned, the VS scheme was originally formulated for the UEF ground state\,\cite{VSsche1}. It is noted that the ground state assumption has strong implications for the self-consistent dielectric formalism itself; when $T=0$ there are no hyperbolic cotangent poles in the frequency-integrated fluctuation--dissipation theorem, the bosonic Matsubara frequencies collapse and the Matsubara infinite sum is substituted with a positive frequency integral. The first rigorous attempt to extend the VS scheme to the finite temperature UEF was due to Sjostrom and Dufty\,\cite{VSsche2}. In what follows, we shall refer to it as VS-SD. Unfortunately, when progressing from the general formulation to the finite temperature formulation, the authors did not take into account the density dependence of the degeneracy parameter $\Theta$ in both the VS SLFC and the CSR expression, see their Eq.(15). As a consequence, the VS-SD SLFC reads as
\begin{align}
G_{\mathrm{VS}}^{\mathrm{SD}}(x;r_{\mathrm{s}},\Theta)&=\left[1+\alpha(r_{\mathrm{s}},\Theta)\left(-\frac{1}{3}r_{\mathrm{s}}\frac{\partial}{\partial{r}_{\mathrm{s}}}-\frac{1}{3}x\frac{\partial}{\partial{x}}\right)\right]\nonumber\\&\quad\times{G}_{\mathrm{STLS}}(x;r_{\mathrm{s}},\Theta)\,,\label{VSSLFCftSD}
\end{align}
and the VS-SD self-consistency parameter equation that stems from the CSR expression reads as
\begin{align}
\alpha_{\mathrm{SD}}(r_{\mathrm{s}},\Theta)=\frac{\displaystyle\left(2-\frac{1}{6}r_{\mathrm{s}}^2\frac{\partial^2}{\partial{r_{\mathrm{s}}^2}}+\frac{4}{3}r_{\mathrm{s}}\frac{\partial}{\partial{r_{\mathrm{s}}}}\right)\widetilde{f}_{\mathrm{xc}}(r_{\mathrm{s}},\Theta)}{\displaystyle\left(1+\frac{1}{3}r_{\mathrm{s}}\frac{\partial}{\partial{r}_{\mathrm{s}}}\right)\widetilde{u}_{\mathrm{int}}(r_{\mathrm{s}},\Theta)}.\label{selfconsistencyVSSD}
\end{align}
It is straightforward that the VS-SD formulation can be directly obtained from the correct VS formulation by setting $\Theta=0$ whenever $\Theta$ appears explicitly, leaving the implicit $\Theta-$dependence of thermodynamic and structural properties intact. It is also apparent that the VS-SD finite temperature formulation results in the same SLFC closure and the same self-consistency parameter equation as the original VS ground state formulation. Nevertheless, the dielectric formalism machinery and thus the numerical solution of the two formulations are different. In spite of the thermodynamic error of the Sjostrom and Dufty analysis, the VS-SD finite temperature scheme is still valuable, since it is numerically much simpler as well as computationally much less costly than the correct VS finite temperature scheme but also since the two schemes should lead to identical predictions at the limit of high degeneracy and / or strong coupling. To sum up, in normalized units, the SD finite temperature formulation of the VS closure consists of Eqs.(\ref{STLSSLFCftcorr},\ref{VSSLFCftSD}) for the SLFC and Eq.(\ref{selfconsistencyVSSD}) for the CSR.

\section{Computational details}

\subsection{Set of equations}\label{subsec:numnormalizedset}

\noindent The closed normalized set of equations emerges by combining the general ingredients of the dielectric formalism with the specific VS closure. The VS equations feature: \textbf{(1)} The normalization condition of the Fermi-Dirac energy distribution function that allows the computation of the reduced chemical potential $\bar{\mu}=\beta\mu$ [see Eq.(\ref{VSfinal1})]. \textbf{(2)} The analytic continuation of the ideal Lindhard density response evaluated at the imaginary Matsubara frequencies $\omega_l=2\pi{l}/(\beta\hbar)$. In particular, an auxiliary complex function $\Phi(\boldsymbol{k},z)=-(2E_{\mathrm{f}}/3n)\widetilde{\chi}_0(\boldsymbol{k},z)$ is introduced which directly leads to $\widetilde{\chi}_0(x,l)/(n\beta)=-(3/2)\Theta\Phi(x,l)$ in the normalized wavenumber Matsubara space\,\cite{TanIchi}. Note that $\Phi(x,l\neq0)$ and $\Phi(x,0)$ are considered separately due to a logarithmic singularity (at $y=x/2$ when $l=0$) that is removable after integration by parts [see Eqs.(\ref{VSfinal2},\ref{VSfinal3})]. \textbf{(3)} The VS SLFC as extensively discussed in Sec.\ref{subsec:FTVS} [see Eq.(\ref{VSfinal4})]. \textbf{(4)} The Matsubara summation expression for the SSF [see Eq.(\ref{VSfinal5})]. In practice, in the numerical implementation, the auxiliary complex function $\Phi(x,l)$ and the square of the combined high frequency large wavenumber asymptotic expansion of the square of the auxiliary complex function $\Phi^2(x,l\to\infty)$ are split up from the infinite series coefficients. Then, the resulting infinite sums can be evaluated exactly leading to the non-interacting (Hartree-Fock) SSF $S_{\mathrm{HF}}(x)$ and a residual SSF correction $S_{\infty}(x)$\,\cite{TanIchi,qSTLSFT,TanConv}. As a direct consequence, there is a significant acceleration of the convergence rate of the Matsubara summation even at strong degeneracy\,\cite{IETChem,IETLett,qIETLet}. \textbf{(5)} The non-linear equation for the theoretical value of the self-consistency parameter $a_{\mathrm{th}}$, \emph{i.e.} the value that exactly satisfies the CSR, as extensively discussed in Sec.\ref{subsec:FTVS} [see Eq.(\ref{VSfinal6})].
\begin{widetext}
\begin{align}
&\int_0^{\infty}\frac{\sqrt{z}dz}{\exp{\left(z-\bar{\mu}\right)}+1}=\frac{2}{3}\Theta^{-3/2}\,,\label{VSfinal1}\\
&\Phi(x,l)=\frac{1}{2x}\int_0^{\infty}\frac{y}{\exp{\left(\frac{y^2}{\Theta}-\bar{\mu}\right)}+1}\ln{\left[\frac{\left(x^2+2xy\right)^2+\left(2\pi{l}{\Theta}\right)^2}{\left(x^2-2xy\right)^2+\left(2\pi{l}\Theta\right)^2}\right]}dy\,,\label{VSfinal2}\\
&\Phi(x,0)=\frac{1}{\Theta{x}}\int_0^{\infty}\frac{y\exp{\left(\frac{y^2}{\Theta}-\bar{\mu}\right)}}{\left[\exp{\left(\frac{y^2}{\Theta}-\bar{\mu}\right)}+1\right]^2}\left[\left(y^2-\frac{x^2}{4}\right)\ln{\left|\frac{2y+x}{2y-x}\right|}+xy\right]dy\,,\label{VSfinal3}\\
&G_{\mathrm{VS}}(x)=\left[1+\alpha\left(-\frac{2}{3}\Theta\frac{\partial}{\partial\Theta}-\frac{1}{3}r_{\mathrm{s}}\frac{\partial}{\partial{r}_{\mathrm{s}}}-\frac{1}{3}x\frac{\partial}{\partial{x}}\right)\right]\left[-\frac{3}{4}\int_0^{\infty}dyy^2\left[S(y)-1\right]\left[1+\frac{x^2-y^2}{2xy}\ln{\left|\frac{x+y}{x-y}\right|}\right]\right]\,,\label{VSfinal4}\\
&S(x)=\frac{3}{2}\Theta\sum_{l=-\infty}^{+\infty}\frac{\Phi(x,l)}{1+\displaystyle\frac{4}{\pi}\lambda{r}_{\mathrm{s}}\frac{1}{x^2}[1-G_{\mathrm{VS}}(x)]\Phi(x,l)}\,,\label{VSfinal5}\\
&\alpha_{\mathrm{th}}=\frac{\displaystyle\left(2-\frac{2}{3}\Theta^2\frac{\partial^2}{\partial{\Theta}^2}-\frac{1}{6}r_{\mathrm{s}}^2\frac{\partial^2}{\partial{r_{\mathrm{s}}^2}}-\frac{2}{3}\Theta{r}_{\mathrm{s}}\frac{\partial^2}{\partial{\Theta}\partial{r_{\mathrm{s}}}}+\frac{1}{3}\Theta\frac{\partial}{\partial{\Theta}}+\frac{4}{3}r_{\mathrm{s}}\frac{\partial}{\partial{r_{\mathrm{s}}}}\right)\widetilde{f}_{\mathrm{xc}}(r_{\mathrm{s}},\Theta)}{\displaystyle\left(1+\frac{2}{3}\Theta\frac{\partial}{\partial\Theta}+\frac{1}{3}r_{\mathrm{s}}\frac{\partial}{\partial{r}_{\mathrm{s}}}\right)\widetilde{u}_{\mathrm{int}}(r_{\mathrm{s}},\Theta)}\,.\label{VSfinal6}
\end{align}
\end{widetext}

\subsection{Numerical algorithm}\label{subsec:numalgorithm}

\noindent The closed normalized set of equations is solved in a home-made hybrid code, where C++ is used for the backend (where all the numerics are handled) and python is used for the frontend (where all the user inputs are introduced, all scheme outputs are extracted and the primary data post-processing is done).

The \emph{first step} concerns the computation of the reduced chemical potential $\bar{\mu}(\Theta)$ for each value of the degeneracy parameter. The solution of Eq.(\ref{VSfinal1}) is found by applying the Brent-Dekker hybrid root-finding algorithm. The \emph{second step} concerns the computation of the ideal Lindhard density response for all Matsubara frequencies from Eqs.(\ref{VSfinal2},\ref{VSfinal3}). For state points close to the ground state, $\Theta\lesssim1$, $l=500$ Matsubara frequencies are required for truncated series convergence. For larger values of the degeneracy parameter, the number of Matsubara frequencies can be safely lowered down to $l=300$. The improper integrals are handled with the doubly-adaptive general-purpose quadrature routine CQUAD of the GSL library; a $0.1$ grid resolution and $50$ upper cut-off are employed.

\begin{figure}
\centering
\resizebox{8.5cm}{!}{\begin{tikzpicture}[x=0.95cm,y=0.95cm,font=\large]
\foreach \Point in {(5.5,0)} {\node at \Point {$+\Delta \Theta$};} 
\foreach \Point in {(0,4.4)} {\node at \Point {$+\Delta r_{\mathrm{s}}$};} 
\foreach \Point in {(0,-5.0)} {\node at \Point {$-\Delta r_{\mathrm{s}}$};}
\foreach \Point in {(-5.7,0)} {\node at \Point {$-\Delta \Theta$};} 
\foreach \Point in {(-3.15,3.14)}{ \node at \Point {\textbullet};}
\foreach \Point in {(-3.5,3.5)}{ \node at \Point {0};}
\foreach \Point in {(-3.0,2.4)}{\node at \Point { $(r_{\mathrm{s}}+\Delta r_{\mathrm{s}},\Theta - \Delta \Theta)$};}
\foreach \Point in {(0,3.13)}{ \node at \Point {\textbullet};}
\foreach \Point in {(-0.5,3.5)}{ \node at \Point {1};}
\foreach \Point in {(0.12,2.4)}{\node at \Point { $(r_{\mathrm{s}}+\Delta r_{\mathrm{s}},\Theta )$};}
\foreach \Point in {(3.15,3.12)}{ \node at \Point {\textbullet};}
\foreach \Point in {(2.8,3.5)}{ \node at \Point {2};}
\foreach \Point in {(3.3,2.4)}{\node at \Point { $(r_{\mathrm{s}}+\Delta r_{\mathrm{s}},\Theta + \Delta \Theta )$};}
\foreach \Point in {(-3.16,-0.04)}{ \node at \Point {\textbullet};}
\foreach \Point in {(-3.5,0.5)}{ \node at \Point {3};}
\foreach \Point in {(-3,-0.8)}{\node at \Point { $(r_{\mathrm{s}}, \Theta- \Delta \Theta )$};}
\foreach \Point in {(-0,-0.04)}{ \node at \Point {\textbullet};}
\foreach \Point in {(-0.5,0.5)}{ \node at \Point {4};}
\foreach \Point in {(0,-0.8)}{\node at \Point { $(r_{\mathrm{s}}, \Theta )$};}
\foreach \Point in {(3.15,-0.05)}{ \node at \Point {\textbullet};}
\foreach \Point in {(2.8,0.5)}{ \node at \Point {5};}
\foreach \Point in {(3.1,-0.8)}{\node at \Point { $(r_{\mathrm{s}}, \Theta + \Delta \Theta)$};}
\foreach \Point in {(-3.16,-3.18)}{ \node at \Point {\textbullet};}
\foreach \Point in {(-3.5,-2.6)}{ \node at \Point {6};}
\foreach \Point in {(-3,-4.0)}{\node at \Point { $(r_{\mathrm{s}}-\Delta r_{\mathrm{s}}, \Theta - \Delta \Theta)$};}
\foreach \Point in {(0,-3.18)}{ \node at \Point {\textbullet};}
\foreach \Point in {(-0.5,-2.6)}{ \node at \Point {7};}
\foreach \Point in {(0.12,-4.0)}{\node at \Point { $(r_{\mathrm{s}}-\Delta r_{\mathrm{s}}, \Theta )$};}
\foreach \Point in {(3.15,-3.18)}{ \node at \Point {\textbullet};}
\foreach \Point in {(2.8,-2.6)}{ \node at \Point {8};}
\foreach \Point in {(3.3,-4.0)}{\node at \Point { $(r_{\mathrm{s}}-\Delta r_{\mathrm{s}}, \Theta + \Delta \Theta)$};}
\draw [dotted, gray] (-5,-4.5) grid (5,4);
\end{tikzpicture}}
\caption{2D grid for the finite difference approximation of the state point derivatives.}\label{fig:computationalstencil}
\end{figure}
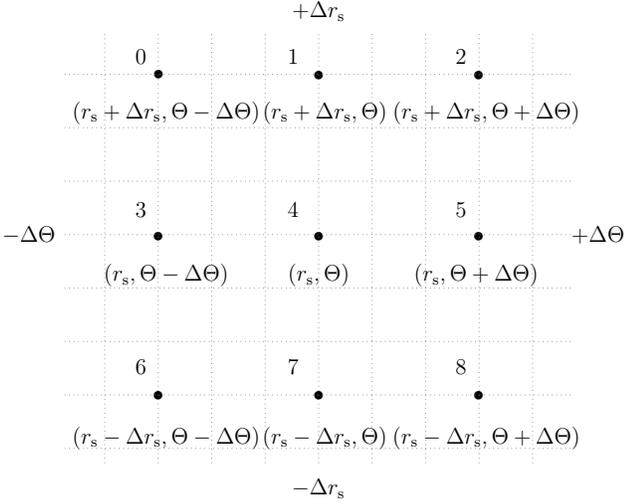

The \emph{third step} concerns the computation of the VS SLFC from Eq.(\ref{VSfinal4}). The STLS SSF is used as the initial value for the SSF. Two initial values are then required for the self-consistency parameter, $\alpha^{(k)}_{\mathrm{num}},\alpha^{(k+1)}_{\mathrm{num}}\in [-0.5,1]$ with $\alpha^{(k)}_{\mathrm{num}}<\alpha^{(k+1)}_{\mathrm{num}}$. As a practical rule, the lower the quantum coupling parameter is, the lower the chosen $\alpha_{\mathrm{num}}$ initial values should be. The reason behind the necessity for computations for two initial values will become apparent when $\alpha_{\mathrm{th}}$ is computed in the fifth step upon the enforcement of the CSR. Moreover, the derivatives with respect to the state point variables $(r_{\mathrm{s}},\Theta)$ need to be computed. Therefore, as deduced from the computational stencil that is shown in Fig.\ref{fig:computationalstencil}, the VS scheme needs to be solved simultaneously for the nine state points $(r_{\mathrm{s}},\Theta),(r_{\mathrm{s}}+\Delta{r}_{\mathrm{s}},\Theta),(r_{\mathrm{s}}-\Delta{r}_{\mathrm{s}},\Theta),(r_{\mathrm{s}},\Theta+\Delta\Theta),(r_{\mathrm{s}},\Theta-\Delta\Theta),(r_{\mathrm{s}}+\Delta{r}_{\mathrm{s}},\Theta+\Delta\Theta),(r_{\mathrm{s}}+\Delta{r}_{\mathrm{s}},\Theta-\Delta\Theta),(r_{\mathrm{s}}-\Delta{r}_{\mathrm{s}},\Theta+\Delta\Theta),(r_{\mathrm{s}}-\Delta{r}_{\mathrm{s}},\Theta-\Delta\Theta)$. The combination of central, forward and backward difference approximations is such that there is no need to supply boundary conditions outside the stencil in order to close the system of equations. In addition, the $x$-derivatives are just computed on a grid depending on how fine the normalized wavevector discretization is. To be more specific, we typically have $\Delta\Theta=0.1$, $\Delta{r}_{\mathrm{s}}=0.1$ and $\Delta{x}=0.1$. With this method, the different state points can be easily stored in a two-dimensional matrix $G[i][j]$ with the $x$ values stored as the $i$-components and the $(r_{\mathrm{s}},\Theta)$ grid as the $j-$components. Since all computations are done up to $x_{\mathrm{max}}=50$, $i$ corresponds to 500 and $j$ to 9. For the $r_{\mathrm{s}}$ state points, which are $(r_{\mathrm{s}},\Theta),(r_{\mathrm{s}},\Theta +\Delta\Theta),(r_{\mathrm{s}},\Theta-\Delta\Theta)$ and with $j=\{3,4,5\}$, the central second order difference is used:
\begin{equation}
\frac{\partial{G}(x;r_{\mathrm{s}})}{\partial{r}_{\mathrm{s}}}\approx\frac{G[i][j+1]-G[i][j-1]}{2\Delta{r}_{\mathrm{s}}}\,.\nonumber
\end{equation}
For the $r_{\mathrm{s}}+\Delta{r}_{\mathrm{s}}$ state points, which are $(r_{\mathrm{s}}+\Delta{r}_{\mathrm{s}},\Theta),(r_{\mathrm{s}}+\Delta{r}_{\mathrm{s}},\Theta+\Delta\Theta),(r_{\mathrm{s}}+\Delta{r}_{\mathrm{s}},\Theta-\Delta\Theta)$ and with $j=\{0,1,2\}$, the forward second order difference is used:
\begin{equation}
\frac{\partial{G}(x;r_{\mathrm{s}}+\Delta{r}_{\mathrm{s}})}{\partial{r}_{\mathrm{s}}}\approx\frac{-3G[i][j]+4G[i][j+1]-G[i][j+2]}{2\Delta{r}_{\mathrm{s}}}\,.\nonumber
\end{equation}
For the $r_{\mathrm{s}}-\Delta{r}_{\mathrm{s}}$ state points, which are $(r_{\mathrm{s}}-\Delta{r}_{\mathrm{s}},\Theta),(r_{\mathrm{s}}-\Delta{r}_{\mathrm{s}},\Theta+\Delta\Theta),(r_{\mathrm{s}}-\Delta{r}_{\mathrm{s}},\Theta-\Delta\Theta)$ and with $j=\{6,7,8\}$, the backward second order difference is used:
\begin{equation}
\frac{\partial{G}(x;r_{\mathrm{s}}-\Delta{r}_{\mathrm{s}})}{\partial{r}_{\mathrm{s}}}\approx\frac{3G[i][j]-4G[i][j-1]+G[i][j-2]}{2\Delta{r}_{\mathrm{s}}}\,.\nonumber
\end{equation}
The same is done for the $\Theta$ derivatives, but now we have $j=\{1,4,7\}$ for the central difference, $j=\{0,3,6\}$ for the forward difference and $j=\{2,5,8\}$ for the backward difference. Finally, the derivatives with respect to $x$ are computed with the central second order difference:
\begin{equation}
\frac{\partial{G}(x)}{\partial{x}}\approx\frac{G[i+1][j]-G[i-1][j]}{2\Delta{x}}\,.\nonumber
\end{equation}
It is emphasized that all finite difference approximations are second order which implies that the truncation errors are $\mathcal{O}[(\Delta{r}_{\mathrm{s}})^2]$, $\mathcal{O}[(\Delta\Theta)^2]$, $\mathcal{O}[(\Delta{x})^2]$. The combination of central, backward and forward differences was chosen so that the grid extent is minimized.

The \emph{fourth step} concerns the computation of the VS SSF from Eq.(\ref{VSfinal5}) for the nine relevant UEF state points. Then, steps 3 and 4 are repeated until the convergence criterion $\sum_{i}\{G_{\mathrm{VS}}^{(n)}[i][4]-G_{\mathrm{VS}}^{(n-1)}[i][4]\}\leq 10^{-5}$ is satisfied for the central point $(r_{\mathrm{s}},\Theta)$, where $(n),(n-1)$ denote the current and previous iteration values of $G_{\mathrm{VS}}$. From now on, this will be referred to as the \enquote{inner loop}. Typically, the number of iterations required to reach convergence for $G_{\mathrm{VS}}$ per numerical self-consistency parameter per state point is of the order of $\sim 100$.

The \emph{fifth step} concerns the convergence of the self-consistency parameter $\alpha_{\mathrm{num}}$ so that the CSR is automatically enforced. The interaction energy $\widetilde{u}_{\mathrm{int}}$ is computed from the SSF, see Eq.(\ref{interaction_energy}). The exchange correlation free energy $\widetilde{f}_{\mathrm{xc}}$ is computed from the adiabatic connection formula, see Eq.(\ref{adiabaticInt}). The respective wavenumber and coupling parameter integrations are again performed with the doubly-adaptive general-purpose quadrature routine CQUAD. Various thermodynamic derivatives of $\widetilde{u}_{\mathrm{int}}$, $\widetilde{f}_{\mathrm{xc}}$ need to be computed. All the involved derivatives are approximated with central differences. For the first-order derivatives (below for $\widetilde{u}_{\mathrm{int}}$ but also for $\widetilde{f}_{\mathrm{xc}}$), we have
\begin{align}
\frac{\partial\widetilde{u}_{\mathrm{int}}(r_{\mathrm{s}},\Theta)}{\partial{r}_{\mathrm{s}}}&\approx\frac{\widetilde{u}_{\mathrm{int}}(r_{\mathrm{s}}+\Delta{r}_{\mathrm{s}},\Theta)-\widetilde{u}_{\mathrm{int}}(r_{\mathrm{s}}-\Delta r_{\mathrm{s}},\Theta)}{2\Delta{r}_{\mathrm{s}}}\,,\nonumber\\
\frac{\partial\widetilde{u}_{\mathrm{int}}(r_{\mathrm{s}},\Theta)}{\partial\Theta}&\approx\frac{\widetilde{u}_{\mathrm{int}}(r_{\mathrm{s}},\Theta+\Delta\Theta)-\widetilde{u}_{\mathrm{int}}(r_{\mathrm{s}},\Theta-\Delta\Theta)}{2\Delta\Theta}\,.\nonumber
\end{align}
For the second-order derivatives, we have
\begin{equation}
\begin{split}
\frac{\partial^2\widetilde{f}_{\mathrm{xc}}(r_{\mathrm{s}},\Theta)}{\partial{r}_{\mathrm{s}}}\approx&\frac{\widetilde{f}_{\mathrm{xc}}(r_{\mathrm{s}}+\Delta{r}_{\mathrm{s}},\Theta)-2\widetilde{f}_{\mathrm{xc}}(r_{\mathrm{s}},\Theta)+\widetilde{f}_{\mathrm{xc}}(r_{\mathrm{s}}-\Delta{r}_{\mathrm{s}},\Theta)}{2\Delta{r}_{\mathrm{s}}}\,,\nonumber\\
\frac{\partial^2\widetilde{f}_{\mathrm{xc}}(r_{\mathrm{s}},\Theta)}{\partial\Theta}\approx&\frac{\widetilde{f}_{\mathrm{xc}}(r_{\mathrm{s}},\Theta+\Delta\Theta)-2\widetilde{f}_{\mathrm{xc}}(r_{\mathrm{s}},\Theta)+\widetilde{f}_{\mathrm{xc}}(r_{\mathrm{s}},\Theta-\Delta\Theta)}{2\Delta\Theta}\,,\nonumber\\
\frac{\partial^2\widetilde{f}_{\mathrm{xc}}(r_{\mathrm{s}},\Theta)}{\partial{r}_{\mathrm{s}}\partial\Theta}\approx&\frac{\widetilde{f}_{\mathrm{xc}}(r_{\mathrm{s}}+\Delta{r}_{\mathrm{s}},\Theta+\Delta\Theta)-\widetilde{f}_{\mathrm{xc}}(r_{\mathrm{s}}+\Delta{r}_{\mathrm{s}},\Theta-\Delta\Theta)}{4\Delta{r}_{\mathrm{s}}\Delta\Theta}-\nonumber\\&\frac{\widetilde{f}_{\mathrm{xc}}(r_{\mathrm{s}}-\Delta{r}_{\mathrm{s}},\Theta+\Delta\Theta)-\widetilde{f}_{\mathrm{xc}}(r_{\mathrm{s}}-\Delta{r}_{\mathrm{s}},\Theta-\Delta\Theta)}{4\Delta{r}_{\mathrm{s}}\Delta\Theta}\,.\nonumber
\end{split}
\end{equation}
These allow the computation of $\alpha_{\mathrm{th}}$ from Eq.(\ref{VSfinal6}). After convergence of the initial inner loop, the $G_{\mathrm{VS}}^{(k)}, G_{\mathrm{VS}}^{(k+1)}$ that correspond to the guesses $\alpha^{(k)}_{\mathrm{num}},\alpha^{(k+1)}_{\mathrm{num}}$ yield the values $\alpha^{(k)}_{\mathrm{th}}(r_{\mathrm{s}},\Theta),\alpha^{(k+1)}_{\mathrm{th}}(r_{\mathrm{s}},\Theta)$. The root of $f(\alpha_{\mathrm{th}},\alpha_{\mathrm{num}})=\alpha_{\mathrm{th}}(r_{\mathrm{s}},\Theta)-\alpha_{\mathrm{num}}=0$, which yields the $\alpha_{\mathrm{num}}$ that satisfies the CSR near exactly, is found with the secant method. After the inner loop has converged, $f^{(k)}=\alpha^{(k)}_{\mathrm{th}}-\alpha^{(k)}_{\mathrm{num}}$ and $f^{(k+1)}=\alpha^{(k+1)}_{\mathrm{th}}-\alpha^{(k+1)}_{\mathrm{num}}$ are computed. If the convergence criterion $|f^{(k+1)}-f^{(k)}|<10^{-3}$ is not satisfied, then the next guess $\alpha^{(k+2)}_{\mathrm{num}}$ is computed via
\begin{equation}
\alpha^{(k+2)}_{\mathrm{num}}=\alpha^{(k+1)}_{\mathrm{num}}-f^{(k+1)}\frac{\alpha^{(k+1)}_{\mathrm{num}}-\alpha^{(k)}_{\mathrm{num}}}{f^{(k+1)}-f^{(k)}}\nonumber\,.
\end{equation}
At this point, the fifth step, which constitutes the \enquote{outer loop} is concluded, and the inner loop is restarted with initial guesses $\alpha^{(k+2)}_{\mathrm{num}},\alpha^{(k+1)}_{\mathrm{num}}$. For the full VS scheme to be solved, the inner and outer loops must be repeated until both convergence criteria are satisfied. On average, the outer loop requires less than five iterations before convergence is reached. Hence, the home-made VS solver is very fast. Typically, the solution of the VS scheme per UEF state point requires 5 sec on a conventional laptop.

\section{Results}

\noindent In this section, the numerical solutions of our finite temperature VS scheme for thermodynamic quantities and static structural properties (SSF, PCF, SDR, SLFC) will be compared with the predictions of the STLS scheme and the VS-SD scheme.

The VS thermodynamic properties will also be compared with highly accurate warm dense UEF equations of state\,\cite{DornRev,EOSdis1,EOSdis2,EOSdis3,EOSdis4}. The GDSMFB parametrization of the UEF exchange correlation free energy $\widetilde{f}_{\mathrm{xc}}$ will be preferred\,\cite{DornRev,EOSdis1}. The GDSMFB parametrization is based on the Ichimaru-Iyetomi-Tanaka (IIT) functional form for $\widetilde{u}_{\mathrm{int}}$\,\cite{IchiRep,EOSdis1,EOSdis5}. The classical analytical Debye-H{\"u}ckel limit ($\Theta\to\infty$)\,\cite{cOCPrev} as well as the Perrot and Dharma-wardana closed-form approximation of the Hartree-Fock limit ($r_{\mathrm{s}}\to0$)\,\cite{EOSdis6} are exactly incorporated. The GDSMFB parametrization primarily utilizes finite-size corrected QMC results\,\cite{newQMCa} obtained by various novel PIMC methods within $0.1\leq{r}_{\mathrm{s}}\leq20$ and $0.5\leq\Theta\leq8$ which are complemented by ground state QMC results within $0.5\leq{r}_{\mathrm{s}}\leq20$\,\cite{EOSdis8}. The GDSMFB parametrization also utilizes synthetic data in the range $0.0625\leq\Theta\leq0.25$ (inaccessible to simulations owing to the prevalence of the fermion sign problem\,\cite{newQMC6}) which are constructed by combining the ground state QMC results with a small STLS-based temperature correction. For the paramagnetic case, the quasi-exact QMC results employed for the fit concern $58$ warm dense UEF state points and $7$ ground state UEF state points\,\cite{EOSdis1,newQMCa,EOSdis8}.

VS-generated structural properties will also be compared with the highly accurate results of the effective static approximation (ESA). The ESA approach is a recently developed semi-empirical scheme of the dielectric formalism that targets the WDM regime (strict applicability limits within $0.7\leq{r}_{\mathrm{s}}\leq20$ and $0\leq\Theta\leq4$)\,\cite{ESApap1,ESApap2}. More specifically, the ESA constructs an analytical SLFC based on the exact long wavelength limit as determined from the CSR\,\cite{ESApap3}, on the exact large wavenumber limit as determined from the cusp relation within the static assumption\,\cite{ESApap4} and on a neural net representation of QMC results at the intermediate wavenumbers\,\cite{ESApap5}. The quasi-exact incorporation of the asymptotic limits is facilitated by the availability of the aforementioned GDSMFB $f_{\mathrm{xc}}$ parametrization and the availability of an on-top PCF (value at zero distance) parametrization\,\cite{ESApap1,EOSdis8,ESApap6}. ESA yields remarkable results for most physical quantities in the WDM regime, not only static (such as the SSF and the electronically screened ion potential) but even dynamic (such as the dynamic structure factor and the stopping power)\,\cite{ESApap2}. In addition, the computational cost of the ESA scheme is negligible, since it is comparable to that of the RPA scheme owing to the fact that the SLFC is introduced as function of the wavenumber and not as a functional of the SSF. Nevertheless, given its QMC-based construction, the ESA scheme cannot provide microscopic insights on the interplay between thermal excitations, strong correlations and quantum effects.

\subsection{Self-consistency parameter}\label{subsec:alpharesults}

\begin{figure}
	\centering
	\includegraphics[width=3.5in]{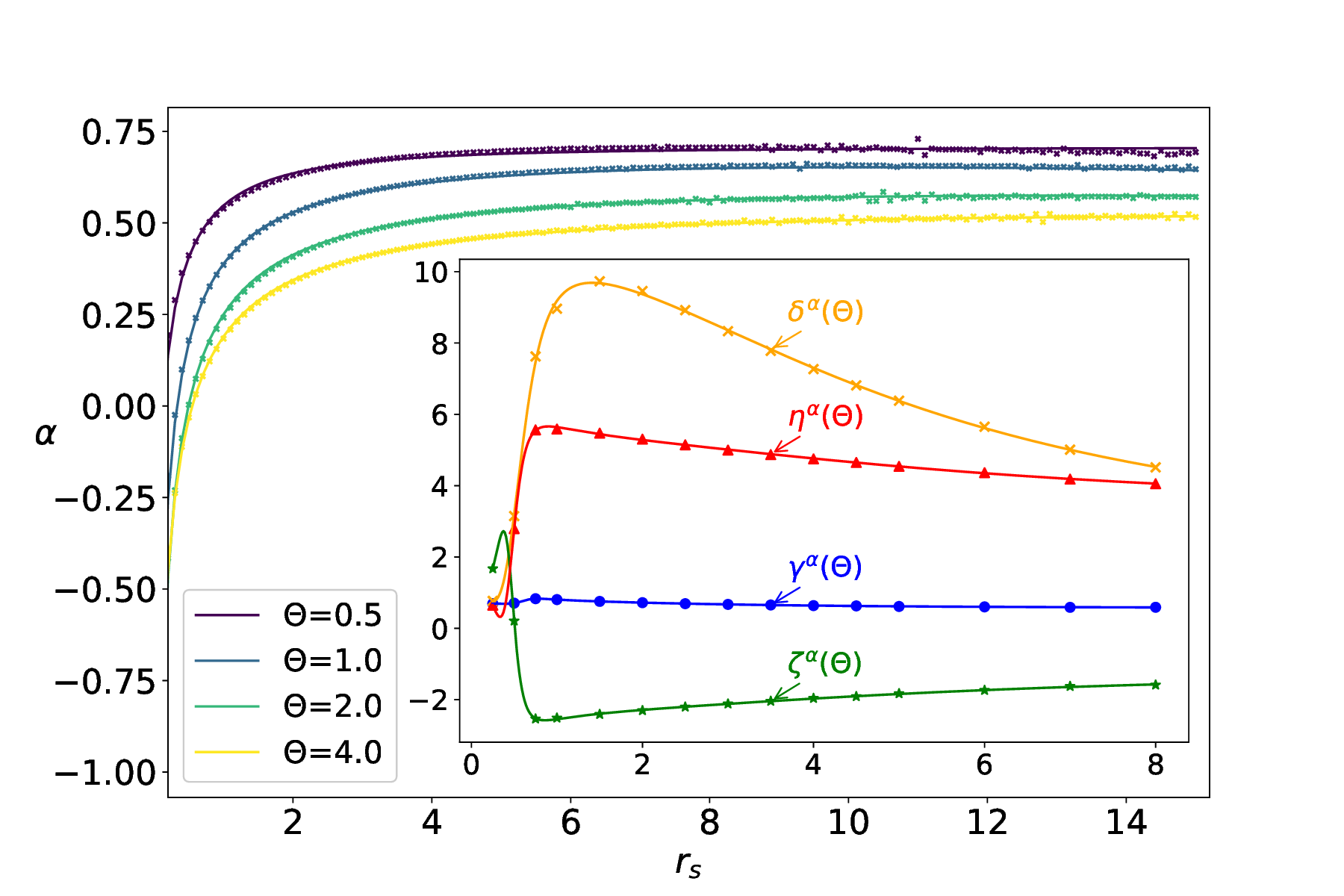}
	\caption{(Main) Dependence of the self-consistency parameter $\alpha$ of the VS scheme on the UEF thermodynamic variables. Numerical data (symbols) and analytic fits (solid lines) for $r_{\mathrm{s}}=0.5-15$ and $\Theta=0.5$ (purple), $\Theta=1.0$ (gray), $\Theta=2.0$ (green), $\Theta=4.0$ (yellow). (Inset) The $\Theta-$dependent functions of the exponential fit. Numerical data (symbols) and Pad{\'e} approximants (solid lines).}\label{alphaself}
\end{figure}

\noindent The complications in the numerical solution of the finite temperature VS scheme primarily originate from the necessity for a double convergence of the inner and outer loops. The availability of a closed-form expression for the self-consistency parameter $\alpha(r_{\mathrm{s}},\Theta)$ is equivalent to cancelling the outer loop and considerably simplifies the VS algorithm. The dependence of the self-consistency parameter $\alpha$ on the UEF thermodynamic variables $(r_{\mathrm{s}},\Theta)$ is explored in Fig.\ref{alphaself}. The following conclusions are due: For any $\Theta$, $\alpha$ is a monotonically increasing function of the quantum coupling parameter that rises abruptly until it reaches an extended plateau. For any $r_{\mathrm{s}}$, $\alpha$ is a monotonically decreasing function of the degeneracy parameter up to $\Theta\sim4$, with the dependence becoming non-monotonic at higher temperatures. In contrast to the ground state limit, $\alpha$ is not bound within $1/2\leq\alpha\lesssim1$\,\cite{VSsche1} and even obtains negative values at very high densities. In contrast to the ground state limit, $\alpha\simeq2/3$\,\cite{VSsche1} does not constitute an accurate approximation, unless $\Theta\lesssim0.5$ and $r_{\mathrm{s}}\gtrsim1$.

The analytic expression for $\alpha(r_{\mathrm{s}},\Theta)$ is obtained by solving the correct finite temperature VS scheme for $2250$ UEF state points spanning the entire warm dense matter regime. To be more specific, $150$ values of the quantum coupling parameter, $r_{\mathrm{s}}\in[0.1,15]$ with a step of $0.1$, are combined with 15 values of the degeneracy parameter $\Theta\in[0.25,0.5,0.75,1,1.5,2,2.5,3,3.5,4,4.5,5,6,7,8]$. In order to take advantage of the monotonic $r_{\mathrm{s}}$ dependence, the isothermal values of $\alpha$ are first least square fitted to
\begin{equation}
\alpha(r_{\mathrm{s}},\Theta)=\gamma^{\alpha}(\Theta)-\delta^{\alpha}(\Theta)e^{-\left[\zeta^{\alpha}(\Theta)r_{\mathrm{s}}^{1/2}+\eta^{\alpha}(\Theta)r_{\mathrm{s}}^{1/3}\right]}\,.\label{alphafitgen}
\end{equation}
The resulting $15$ data points for the four $\Theta-$dependent functions $\kappa^{\alpha}=\{\gamma^{\alpha},\delta^{\alpha},\zeta^{\alpha},\eta^{\alpha}\}$ are then fitted to even Pad{\'e} approximants of the order 6/6,
\begin{equation}
\kappa^{\alpha}(\Theta)=\frac{\kappa^{\alpha}_1+\kappa^{\alpha}_2\Theta^2+\kappa^{\alpha}_3\Theta^4+\kappa^{\alpha}_4\Theta^6}{1+\kappa^{\alpha}_5\Theta^2+\kappa^{\alpha}_6\Theta^4+\kappa^{\alpha}_7\Theta^6}\,.\label{alphafitpade}
\end{equation}
The $28$ Pad{\'e} coefficients are provided in Table \ref{selfconsistency_table} together with the mean absolute relative error of each Pad{\'e} fit. The Pad{\'e} approximants are plotted in the inset of Fig.\ref{alphaself}. The overall mean absolute relative error of the total fit is $1.4\%$. The errors stem mostly from the plateau region, where there are weak $\alpha$ fluctuations that are smoothed out by the fitting function, see Fig.\ref{alphaself}.

With the aid of this $\alpha(r_{\mathrm{s}},\Theta)$ expression, the VS scheme can now be solved by algorithms that simultaneously solve the STLS scheme for nine UEF state points. Given the $\sim1\%$ fitting error and the $10^{-3}$ convergence criterion of the outer loop, it can be roughly stated that such STLS algorithms will solve the VS scheme with approximately an order of accuracy less than our pure VS algorithm.

\begin{table}
    \caption{Pad{\'e} coefficients for the self-consistency parameter of the VS scheme, see Eqs.(\ref{alphafitgen},\ref{alphafitpade}). The mean absolute relative error of each Pad{\'e} fit is reported in the last row.}\label{selfconsistency_table}
	\centering
	\begin{tabular}{ccccc}
    \hline
    $i$       & $\gamma_i^{\alpha}$     & $\delta_i^{\alpha}$     &  $\zeta_i^{\alpha}$     &  $\eta_i^{\alpha}$    \\ \hline
    $1$       & \,\,\,$0.7002$\,\,\,    & \,\,\,$1.1498$\,\,\,    & \,\,\,$0.5854$\,\,\,    & \,\,\,$1.5104$\,\,\,  \\
    $2$       & \,\,\,$-2.7913$\,\,\,   & \,\,\,$-12.9044$\,\,\,  & \,\,\,$10.6079$\,\,\,   & \,\,\,$-22.9869$\,\,\,\\
    $3$       & \,\,\,$4.4473$\,\,\,    & \,\,\,$101.674$\,\,\,   & \,\,\,$-49.863$\,\,\,   & \,\,\,$99.745$\,\,\,  \\
    $4$       & \,\,\,$-3.7928$\,\,\,   & \,\,\,$-1.0153$\,\,\,   & \,\,\,$-7.2311$\,\,\,   & \,\,\,$-5.7303$\,\,\, \\
    $5$       & \,\,\,$0.4222$\,\,\,    & \,\,\,$0.8387$\,\,\,    & \,\,\,$-0.8369$\,\,\,   & \,\,\,$1.8335$\,\,\,  \\
    $6$       & \,\,\,$5.4892$\,\,\,    & \,\,\,$9.5142$\,\,\,    & \,\,\,$21.0284$\,\,\,   & \,\,\,$18.3802$\,\,\, \\
    $7$       & \,\,\,$0.7531$\,\,\,    & \,\,\,$0.3872$\,\,\,    & \,\,\,$0.6994$\,\,\,    & \,\,\,$0.5487$\,\,\,  \\
    $\%$      & \,\,\,$0.23$\,\,\,      & \,\,\,$0.71$\,\,\,      & \,\,\,$0.49$\,\,\,      & \,\,\,$0.26$\,\,\,    \\ \hline
	\end{tabular}
\end{table}

\subsection{Exchange correlation free energy}\label{subsec:fxcresults}

\begin{figure}
	\centering
	\includegraphics[width=3.5in]{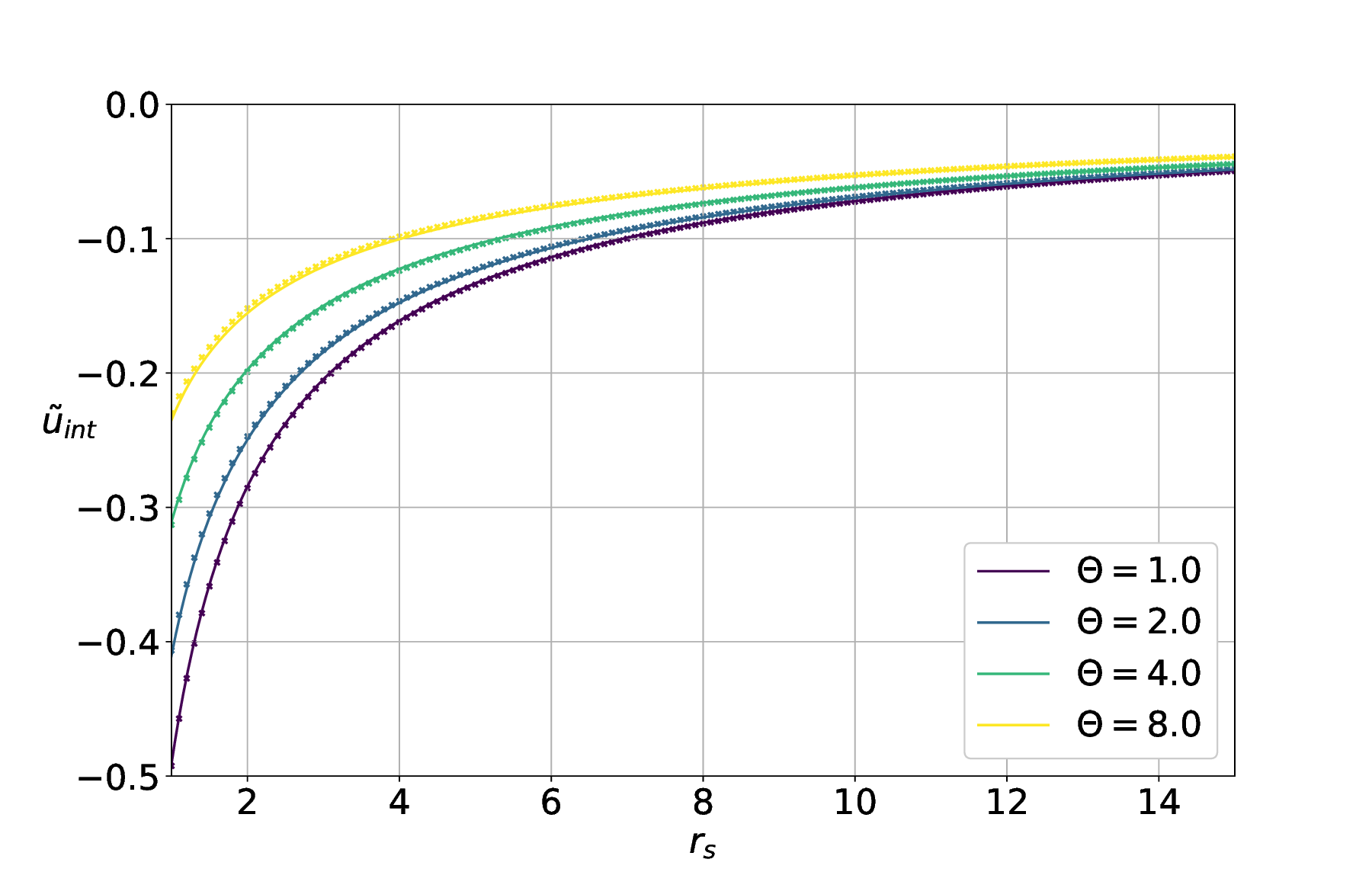}
	\caption{Equation of state for the UEF interaction energy of the VS scheme. Numerical data (symbols) and analytic fits (solid lines) for $r_{\mathrm{s}}=1-15$ and $\Theta=1$ (purple), $\Theta=2$ (blue), $\Theta=4$ (green), $\Theta=8$ (yellow). The $\widetilde{u}_{\mathrm{int}}$ fit follows Eqs.(\ref{fxcfitgen},\ref{fxcfitpade}) but the Pad{\'e} coefficients are not provided in the main text.}\label{EOSuint}
\end{figure}

\begin{figure}
	\centering
	\includegraphics[width=3.5in]{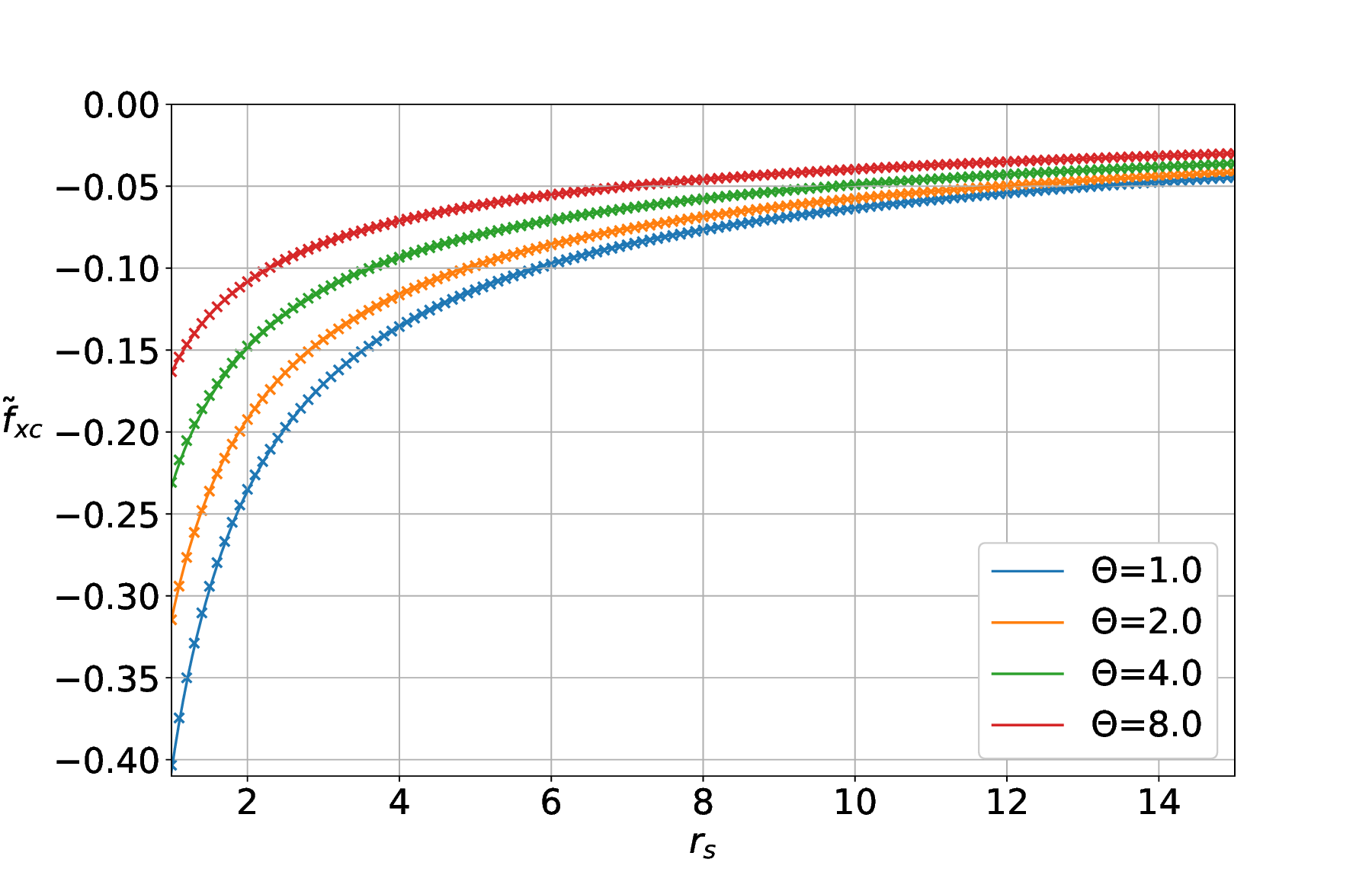}
	\caption{Equation of state for the UEF exchange correlation free energy of the VS scheme. Numerical data (symbols) and analytic fits (solid lines) for $r_{\mathrm{s}}=1-15$ and $\Theta=1.0$ (blue), $\Theta=2.0$ (orange), $\Theta=4.0$ (green), $\Theta=8.0$ (red). The $\widetilde{f}_{\mathrm{{xc}}}$ fit follows Eqs.(\ref{fxcfitgen},\ref{fxcfitpade}) with the Pad{\'e} coefficients of Table \ref{fxc_parametrization_table}.}\label{EOSfxc}
\end{figure}

\noindent As per usual, the VS SSF is employed for the determination of the UEF interaction energy $\widetilde{u}_{\mathrm{int}}$ via the fundamental expression of Eq.(\ref{interaction_energy}). Then, the UEF interaction energy is employed for the computation of the UEF exchange correlation free energy $\widetilde{f}_{\mathrm{xc}}$ via the integral version of the adiabatic connection formula of Eq.(\ref{adiabaticInt}). Different thermodynamic properties can then be computed from well known expressions\,\cite{quanele}. The VS-generated $\widetilde{u}_{\mathrm{int}}$ and $\widetilde{f}_{\mathrm{xc}}$ have the standard UEF dependencies\,\cite{DornRev}, as seen in Figs.\ref{EOSuint},\ref{EOSfxc}. In what follows, we exclusively focus on $\widetilde{f}_{\mathrm{xc}}$.

The $\widetilde{f}_{xc}(r_{\mathrm{s}},\Theta)$ parametrization is based on the solution of the finite temperature VS scheme for $2250$ UEF states that cover the entire WDM regime. Again, $150$ values of the quantum coupling parameter, $r_{\mathrm{s}}\in[0.1,15]$ with a step of $0.1$, and 15 values of the degeneracy parameter $\Theta\in[0.25,0.5,0.75,1,1.5,2,2.5,3,3.5,4,4.5,5,6,7,8]$ are probed. Our $\widetilde{f}_{\mathrm{{xc}}}$ parametrization is inspired by the IIT parametrization originally proposed for $\widetilde{u}_{\mathrm{int}}$\,\cite{DornRev,IchiRep,EOSdis1}. The isothermal values are first least square fitted to
\begin{equation}
\widetilde{f}_{\mathrm{{xc}}}(r_{\mathrm{s}},\Theta)=-\frac{1}{r_{\mathrm{s}}}\frac{\alpha_{\mathrm{HF}}(\Theta)+\gamma^{\mathrm{f}}(\Theta)\sqrt{r_{\mathrm{s}}}+\delta^{\mathrm{f}}(\Theta)r_{\mathrm{s}}}{1+\zeta^{\mathrm{f}}(\Theta)\sqrt{r_{\mathrm{s}}}+\eta^{\mathrm{f}}(\Theta)r_{\mathrm{s}}}\,,\label{fxcfitgen}
\end{equation}
where $\alpha_{\mathrm{HF}}(\Theta)$ corresponds to the exact Hartree-Fock coefficient of the non-interacting UEF exchange free energy $\widetilde{f}(r_{\mathrm{s}},\Theta)=-\alpha_{\mathrm{HF}}(\Theta)/r_{\mathrm{s}}$ that is rigorously defined by the double integral presentation
\begin{widetext}
\begin{align*}
\alpha_{\mathrm{HF}}(\Theta)=\frac{3}{4\pi\lambda}\Theta\int_0^{\infty}\int_0^{\infty}\frac{y}{x}\frac{1}{\exp{\left(\frac{y^2}{\Theta}-\bar{\mu}\right)}+1}\ln{\left|\frac{1+\exp{\left[\bar{\mu}-\frac{(y-x)^2}{\Theta}\right]}}{1+\exp{\left[\bar{\mu}-\frac{(y+x)^2}{\Theta}\right]}}\right|}dydx\,,
\end{align*}
and has been approximated by the very accurate Perrot and Dharma-wardana parametrization\,\cite{EOSdis6}
\begin{align*}
\alpha_{\mathrm{HF}}(\Theta)=\frac{1}{\pi\lambda}\tanh{\left(\frac{1}{\Theta}\right)}\frac{0.75+3.04363\Theta^2-0.09227\Theta^3+1.7035\Theta^4}{1+ 8.31051\Theta^2+5.1105\Theta^4}\,.
\end{align*}
\end{widetext}
The resulting $15$ data points for the four $\Theta-$dependent functions $\kappa^{\mathrm{f}}=\{\gamma^{\mathrm{f}},\delta^{\mathrm{f}},\zeta^{\mathrm{f}},\eta^{\mathrm{f}}\}$ are then fitted to even Pad{\'e} approximants of the order 6/6,
\begin{equation}
\kappa^{\mathrm{f}}(\Theta)=\frac{\kappa^{\mathrm{f}}_1+\kappa^{\mathrm{f}}_2\Theta^2+\kappa^{\mathrm{f}}_3\Theta^4+\kappa^{\mathrm{f}}_4\Theta^6}{1+\kappa^{\mathrm{f}}_5\Theta^2+\kappa^{\mathrm{f}}_6\Theta^4+\kappa^{\mathrm{f}}_7\Theta^6}\,.\label{fxcfitpade}
\end{equation}
The $28$ Pad{\'e} coefficients are provided in Table \ref{fxc_parametrization_table} together with the mean absolute relative error of each Pad{\'e} fit. The overall mean absolute relative error of the total fit is merely $0.36\%$.

\begin{table}
    \caption{Pad{\'e} coefficients for the exchange correlation free energy of the VS scheme, see Eqs.(\ref{fxcfitgen},\ref{fxcfitpade}). The mean absolute relative error of each Pad{\'e} fit is reported in the last row.}\label{fxc_parametrization_table}
	\centering
	\begin{tabular}{ccccc}
    \hline
    $i$       & $\gamma_i^{\mathrm{f}}$ & $\delta_i^{\mathrm{f}}$ &  $\zeta_i^{\mathrm{f}}$ &  $\eta_i^{\mathrm{f}}$          \\ \hline
    $1$       & \,\,\,$43.9016$\,\,\,   & \,\,\,$30.0595$\,\,\,   & \,\,\,$99.9987$\,\,\,   & \,\,\,$35.8655$\,\,\,  \\
    $2$       & \,\,\,$103.113$\,\,\,   & \,\,\,$150.869$\,\,\,   & \,\,\,$ 99.5025$\,\,\,  & \,\,\,$98.756$\,\,\,   \\
    $3$       & \,\,\,$17.751$\,\,\,    & \,\,\,$150.733$\,\,\,   & \,\,\,$71.9001$\,\,\,   & \,\,\,$98.331$\,\,\,   \\
    $4$       & \,\,\,$0.0798$\,\,\,    & \,\,\,$1.9005$\,\,\,    & \,\,\,$101.654$\,\,\,   & \,\,\,$0.5388$\,\,\,   \\
    $5$       & \,\,\,$6.2362$\,\,\,    & \,\,\,$ 1.4741$\,\,\,   & \,\,\,$0.995$\,\,\,     & \,\,\,$ 0.4205$\,\,\,  \\
    $6$       & \,\,\,$4.2003$\,\,\,    & \,\,\,$4.429$\,\,\,     & \,\,\,$0.719$\,\,\,     & \,\,\,$ 3.1564$\,\,\,  \\
    $7$       & \,\,\,$0.2626$\,\,\,    & \,\,\,$0.1749$\,\,\,    & \,\,\,$1.0165$\,\,\,    & \,\,\,$0.1049$\,\,\,   \\
    $\%$      & \,\,\,$0.03$\,\,\,      & \,\,\,$0.05$\,\,\,      & \,\,\,$0.09$\,\,\,      & \,\,\,$0.14$\,\,\,     \\ \hline
	\end{tabular}
\end{table}

\begin{figure}
	\centering
	\includegraphics[width=3.5in]{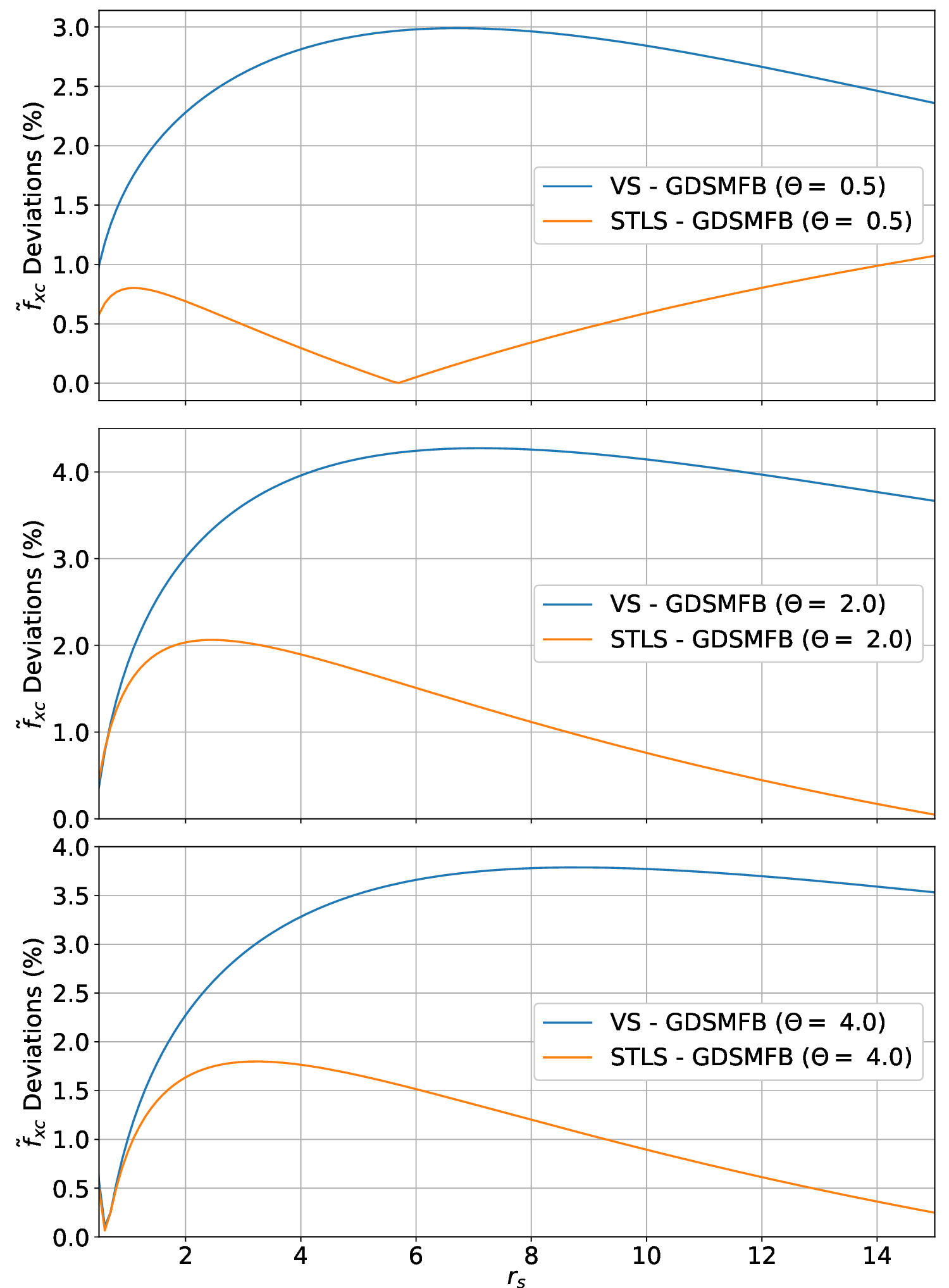}
	\caption{The absolute relative deviations of the VS $\widetilde{f}_{\mathrm{xc}}$ predictions (blue) and the STLS $\widetilde{f}_{\mathrm{xc}}$ predictions (orange) from the very accurate $\widetilde{f}_{\mathrm{xc}}$ value of the GDSMFB parametrization as a function of $r_{\mathrm{s}}$ for $\Theta=0.5$ (top), $\Theta=2.0$ (centre) and $\Theta=4.0$ (bottom). In the case of $\Theta=0.5$, the abrupt monotonicity change for the STLS scheme that occurs at $r_{\mathrm{s}}\simeq5.5$ corresponds to the switch of the sign of the difference.}\label{EOSfxcdev}
\end{figure}

A direct comparison between the VS-generated $\widetilde{f}_{xc}$ and STLS-generated $\widetilde{f}_{xc}$ can be performed neither graphically (small differences) nor tabularly (too many data points). An indirect comparison will be facilitated by the very accurate GDSMFB parametrization. The absolute relative deviations of the VS and STLS $\widetilde{f}_{xc}$ predictions from the GDSMFB $\widetilde{f}_{xc}$ value as a function of $r_{\mathrm{s}}$ have been plotted in Fig.\ref{EOSfxcdev} for $3$ $\Theta$ values. Raw VS and STLS data have been preferred over the analytic fits for the purposes of comparison. Thus, the smoothness of the curves is a direct manifestation of the stability and accuracy of our algorithm. This analysis has been pursued for all $15$ probed $\Theta$. In the entire WDM regime, the STLS predictions $\widetilde{f}_{xc}$ are more accurate. The reason for the thermodynamic superiority of the STLS has been earlier discussed in the literature and will be elucidated in Sec.\ref{subsec:SSFresults}. As we shall see, this does not translate to structural superiority.

\subsection{Static structure factor}\label{subsec:SSFresults}

\begin{figure*}
	\centering
	\includegraphics[width=7.0in]{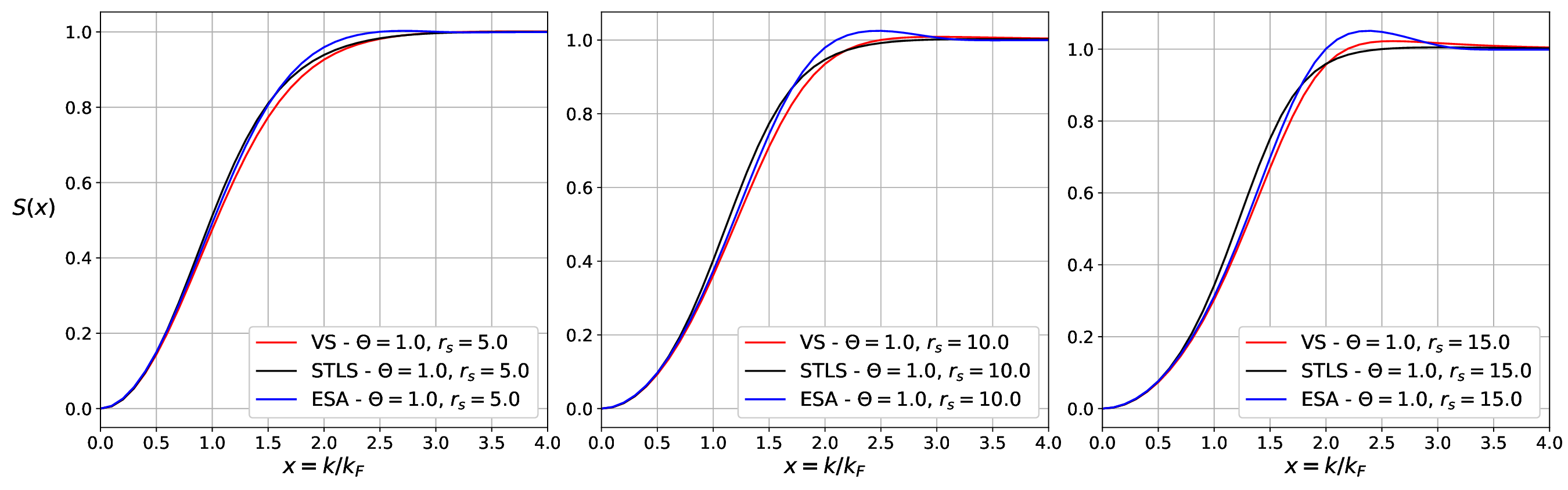}
	\caption{The static structure factor of the paramagnetic uniform electron fluid in the warm dense matter regime, as predicted by the VS scheme (red), the STLS scheme (black) and the quasi-exact ESA scheme (blue). Results for $\Theta=1.0$ and for $r_{\mathrm{s}}=5$ (left), $r_{\mathrm{s}}=10$ (center), $r_{\mathrm{s}}=15$ (right).}\label{SSF_VSvsSTLS}
\end{figure*}

\noindent A comparison between the SSF predictions of the STLS, VS and ESA schemes is featured in Fig.\ref{SSF_VSvsSTLS}. \textbf{(i)} The STLS-generated SSF is consistently characterized by an overestimation at low wavenumbers followed by an underestimation at intermediate and high wavenumbers, with the crossover point lying within $1.5k_{\mathrm{F}}-2.0k_{\mathrm{F}}$. On the other hand, the VS-generated SSF is consistently characterized by an underestimation at low and intermediate wavenumbers followed by a slight overestimation at high wavenumbers, with the crossover point lying in the vicinity of $3k_{\mathrm{F}}$. Since the integrated SSF yields the interaction energy, see Eq.(\ref{interaction_energy}), this observation suggests a favorable cancellation of errors for the STLS scheme and weak cancellation of errors for the VS scheme, which explains the thermodynamic supremacy of the STLS scheme. \textbf{(ii)} At sufficiently high densities for which the SSF maximum is either extremely shallow or even nonexistent, the STLS scheme is consistently more accurate than the VS scheme in nearly the entire wavenumber range. The ESA SSF lies in between the STLS SSF and VS SSF at small wavenumbers, while $S_{\mathrm{ESA}}>S_{\mathrm{STLS}}>S_{\mathrm{VS}}$ is valid at intermediate and large wavenumbers. \textbf{(iii)} At stronger coupling parameters for which the SSF maximum is apparent but remains shallow, the accuracy of the STLS and VS schemes becomes comparable. The ESA-generated SSF is nearly overlapping with the VS SSF at small wavenumbers, in better agreement with the STLS SSF at intermediate wavenumbers and slightly closer to the VS SSF near the large wavenumber limit. It should also be noted that the VS scheme leads to an extremely shallow SSF maximum, while the STLS SSF remains bounded below unity. \textbf{(iv)} Near the boundary between the WDM and the strong coupling regimes where the SSF maximum is pronounced and the secondary extrema begin to emerge, the VS scheme is superior with the exception of a narrow $1.5k_{\mathrm{F}}-2.0k_{\mathrm{F}}$ interval located in the neighborhood of the ESA-STLS crossover. More specifically, the VS-generated SSF is nearly indistinguishable from the ESA-generated SSF within the small wavenumber range and it is characterized by a well-developed maximum, albeit weaker and broader than the nearly exact prediction of the ESA scheme. Meanwhile, the STLS-generated SSF remains bounded below unity.

At strong coupling, as far as static structural properties are concerned, the improved performance of the VS scheme compared to the STLS scheme is expected from a general equilibrium statistical mechanics perspective\,\cite{normal1}. Let us first recall an exact relationship between successive order reduced density matrices which reads as\,\cite{quankin}
\begin{align*}
\varrho_{s}(\boldsymbol{r}^s;\boldsymbol{R}^s)=\frac{1}{N-s}\int\,d^3r_{s+1}\varrho_{s+1}(\boldsymbol{r}^s,\boldsymbol{r}_{s+1};\boldsymbol{R}^s,\boldsymbol{r}_{s+1})\,,
\end{align*}
where $\varrho_{s}(\boldsymbol{r}^s;\boldsymbol{R}^s)=\langle\boldsymbol{r}^s|\hat{\varrho}_s|\boldsymbol{R}^s\rangle$ is the reduced s-particle density matrix within the coordinate representation and $\boldsymbol{x}^s=\{\boldsymbol{x}_1,\boldsymbol{x}_2,\cdots,\boldsymbol{x}_s\}$ for the positions $\boldsymbol{x}=\boldsymbol{r},\boldsymbol{R}$ of the $s\leq{N}$ electrons. In terms of reduced s-particle densities $n_{s}(\boldsymbol{r}^s)=\varrho_{s}(\boldsymbol{r}^s;\boldsymbol{r}^s)$, this relationship remains intact, i.e.,
\begin{align*}
n_{s}(\boldsymbol{r}^s)=\frac{1}{N-s}\int\,d^3r_{s+1}n_{s+1}(\boldsymbol{r}^{s+1})\,.
\end{align*}
Within the thermodynamic limit $s\ll{N}\to\infty$ and for homogeneous systems, in terms of reduced s-particle correlation functions $g_{s}(\boldsymbol{r}^s)=n_{s}(\boldsymbol{r}^s)/n^s$, this relationship becomes\,\cite{normal1,normal2}
\begin{equation}
g_{s}(\boldsymbol{r}^s)=\frac{1}{V}\int\,d^3r_{s+1}g_{s+1}(\boldsymbol{r}^{s+1})\,.\label{ternary_connection}
\end{equation}
Setting $s=2$, the above implies a connection between the (isothermal) density derivative of the pair correlation function and the triplet correlation function. Such connections have been long exploited in the theory of classical liquids\,\cite{normal3,normal4,normal5}. Thus, it is concluded that the indirect inclusion of ternary correlations in the VS scheme is responsible for the improved description of equal-time density correlations compared to the STLS scheme.

\begin{figure}
	\centering
	\includegraphics[width=3.50in]{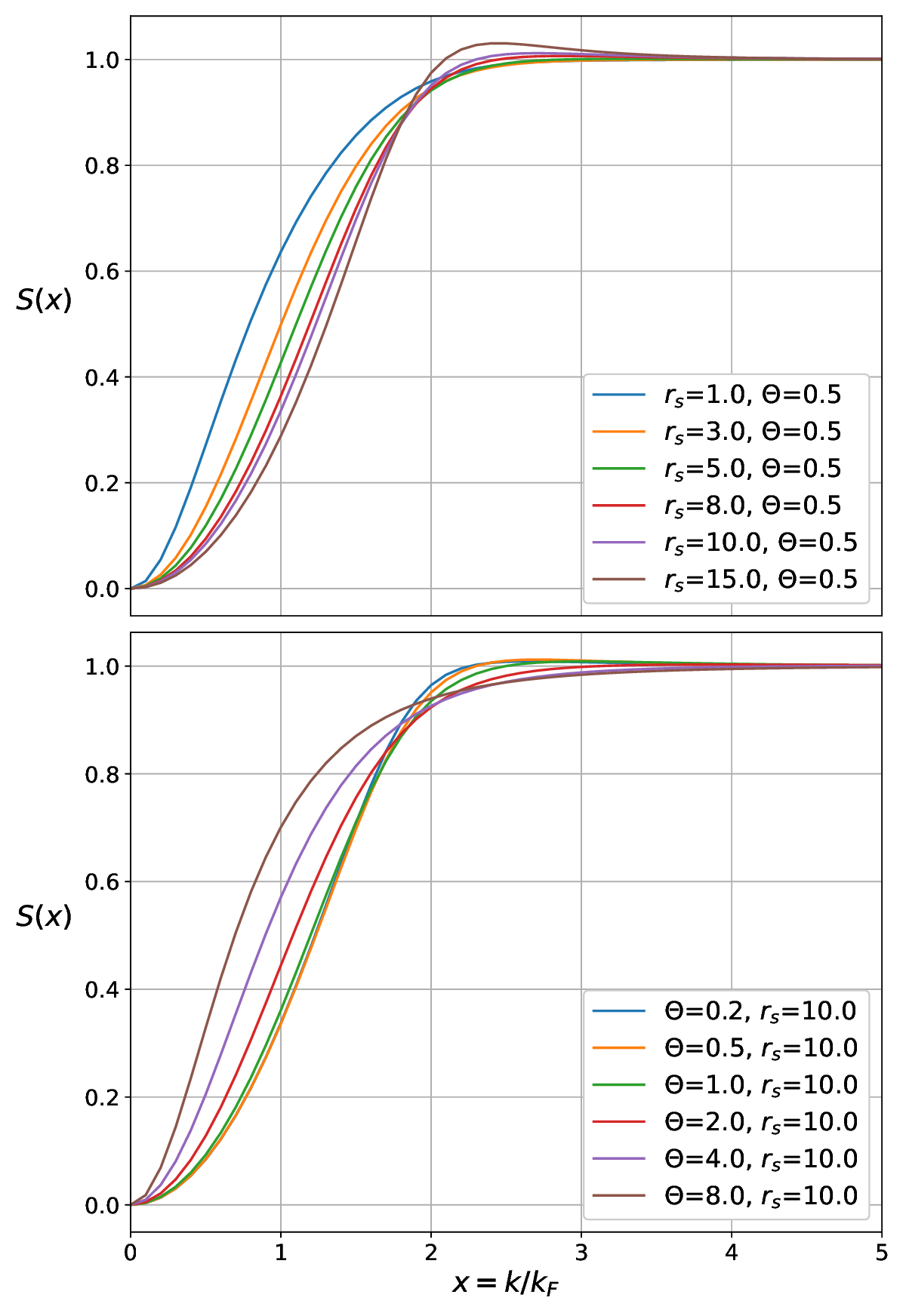}
	\caption{The dependence of the static structure factor of the paramagnetic warm dense uniform electron fluid on the thermodynamic variables $(r_{\mathrm{s}},\Theta)$, as predicted by the finite temperature VS scheme. (Top) Results for constant $\Theta=0.5$ and for varying $r_{\mathrm{s}}=\{1,3,5,8,10,15\}$. (Bottom) Results for constant $r_{\mathrm{s}}=10$ and for varying $\Theta=\{0.2,0.5,1.0,2.0,4.0,8.0\}$.}\label{SSF_VSparametric}
\end{figure}

The dependence of the VS-generated SSFs on the UEF thermodynamic variables $(r_{\mathrm{s}},\Theta)$ is explored in Fig.\ref{SSF_VSparametric}. At constant $\Theta$ and for increasing $r_{\mathrm{s}}$; the small wavenumber behavior of the SSF becomes less steep. When $\Theta=0.5$, a shallow maximum appears for $r_{\mathrm{s}}\lesssim8$ which becomes sharper as the quantum coupling parameter further increases. For all probed isothermal state points, the large wavenumber limit of unity is always reached before $4.5k_{\mathrm{F}}$. Essentially, the $\Theta-$dependency follows the same reasoning as the $r_{\mathrm{s}}-$dependency, provided that an effective coupling parameter $\Gamma_{\mathrm{eff}}$ is introduced which considers the contribution of exchange effects on an equal footing with thermal effects. In view of the classical coupling parameter, one now defines $\Gamma_{\mathrm{eff}}=e^2/(d\sqrt{T^2+E_{\mathrm{f}}^2})$ which yields $\Gamma_{\mathrm{eff}}=2\lambda^2r_{\mathrm{s}}/\sqrt{1+\Theta^2}$\,\cite{BoniRev}. For constant $r_{\mathrm{s}}$ and increasing $\Theta$, the small wavenumber behavior of the SSF becomes steeper. When $r_{\mathrm{s}}=10$, a shallow maximum appears for $\Theta\lesssim2$ that becomes sharper as the degeneracy parameter further decreases. For all probed isochoric state points, the large wavenumber limit of unity is reached before $5k_{\mathrm{F}}$.

\subsection{Static density response function}\label{subsec:SDRresults}

\noindent The linear SDR function is directly obtained from PIMC results for the imaginary time density--density correlation function via the imaginary--time version of the quantum fluctuation--dissipation theorem\,\cite{sdrref1,sdrref2,sdrref3}. The SDR function is known to be more sensitive to quantum effects compared to the SSF\,\cite{sdrref3}. Nevertheless, no surprises are expected in the course of the comparison between the VS scheme and the STLS scheme, given the fact that both schemes treat quantum effects at the level of the RPA. It is pointed out that the large wavenumber limit of the SDR is dictated by the ideal SDR, i.e.,\,\cite{sdrref4,sdrref5}
\begin{equation}
\lim_{k\to\infty}\chi(\boldsymbol{k})=\chi_0(\boldsymbol{k})\,,\label{asymptotic_ideal}
\end{equation}
and that the long wavelength limit of the SDR is dictated by the perfect screening, i.e.\,\cite{sdrref4,sdrref5}
\begin{equation}
\lim_{k\to0}\chi(\boldsymbol{k})=-\frac{k^2}{4\pi{e}^2}\,.\label{asymptotic_perfectscreening}
\end{equation}

\begin{figure*}
	\centering
	\includegraphics[width=7.0in]{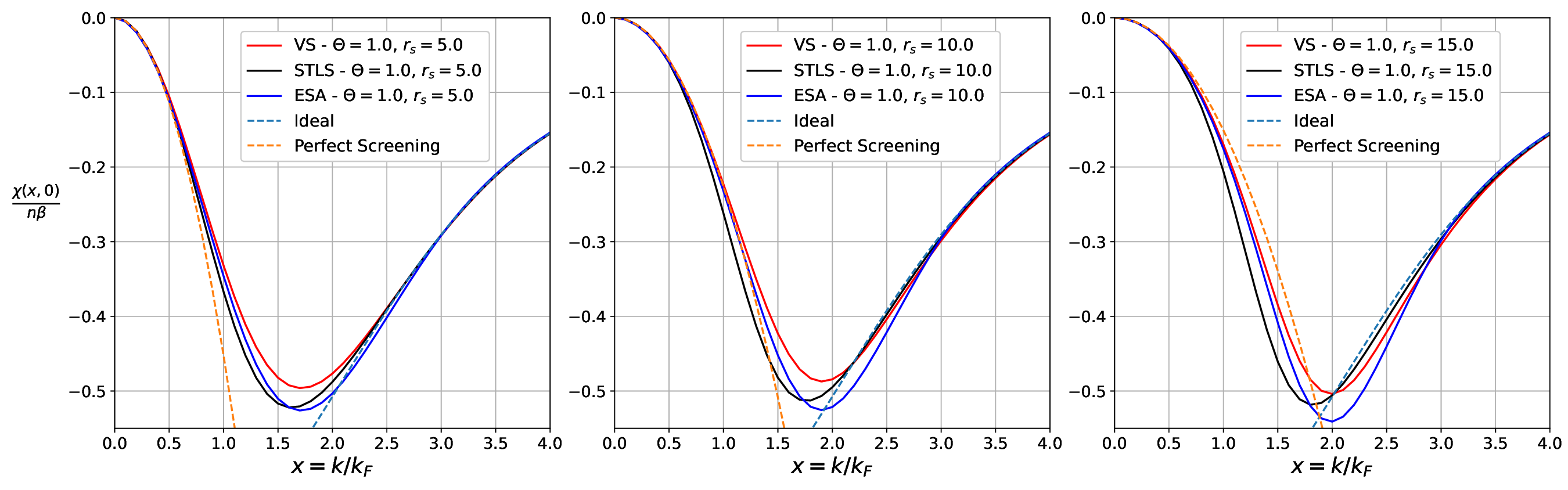}
	\caption{The static density response of the paramagnetic uniform electron fluid in the warm dense matter regime, as predicted by the VS scheme (red solid line), the STLS scheme (black solid line) and the quasi-exact ESA scheme (blue solid line). The ideal non-interacting limit (green dashed line) and the perfect screening limit (orange dashed line) are also indicated. Results for $\Theta=1.0$ and for $r_{\mathrm{s}}=5$ (left), $r_{\mathrm{s}}=10$ (center), $r_{\mathrm{s}}=15$ (right).}\label{SDR_VSvsSTLS}
\end{figure*}

A comparison of the SDR predictions of the STLS, VS and ESA schemes is featured in Fig.\ref{SDR_VSvsSTLS}. The asymptotic limits are indicated by the dashed lines. \textbf{(i)} In accordance with the SSF observations, in the small wavenumber region, the quasi-exact ESA SDR lies in-between the VS-generated and STLS-generated SDRs at high densities. When the quantum coupling parameter $r_{\mathrm{s}}$ increases, the VS-generated SDR nearly overlaps with the ESA SDR for $k\lesssim1.5k_{\mathrm{F}}$, while the STLS-generated SDR exhibits observable deviations from the ESA SDR up to $k/k_{\mathrm{F}}\sim0.7$. It is noted that the perfect screening limit is reached for smaller wavenumbers, as the coupling becomes stronger. \textbf{(ii)} Again in accordance with the SSF observations, in the large wavenumber region, the quasi-exact ESA SDRs lie closer to the VS-generated SDRs irrespective of the quantum coupling parameter. Unsurprisingly, as the coupling becomes stronger, the convergence to the ideal SDR is shifted to higher wavenumbers. \textbf{(iii)} Irrespective of the quantum coupling parameter, the STLS scheme yields more accurate predictions for the magnitude of the SDR minimum, while the VS scheme yields more accurate predictions for the position of the SDR minimum. In particular, as the coupling increases; the STLS predictions for the magnitude of the minimum become progressively worse while the offset concerning the position of the minimum remains nearly constant, the VS predictions for the position of the minimum remain very accurate while the discrepancy concerning the magnitude of the minimum remains nearly constant. \textbf{(iv)} Overall, also in line with the SSF observations, the STLS-generated SDRs are more accurate for weak coupling and the VS-generated SDRs are more accurate for moderate coupling.

\begin{figure}
	\centering
	\includegraphics[width=3.40in]{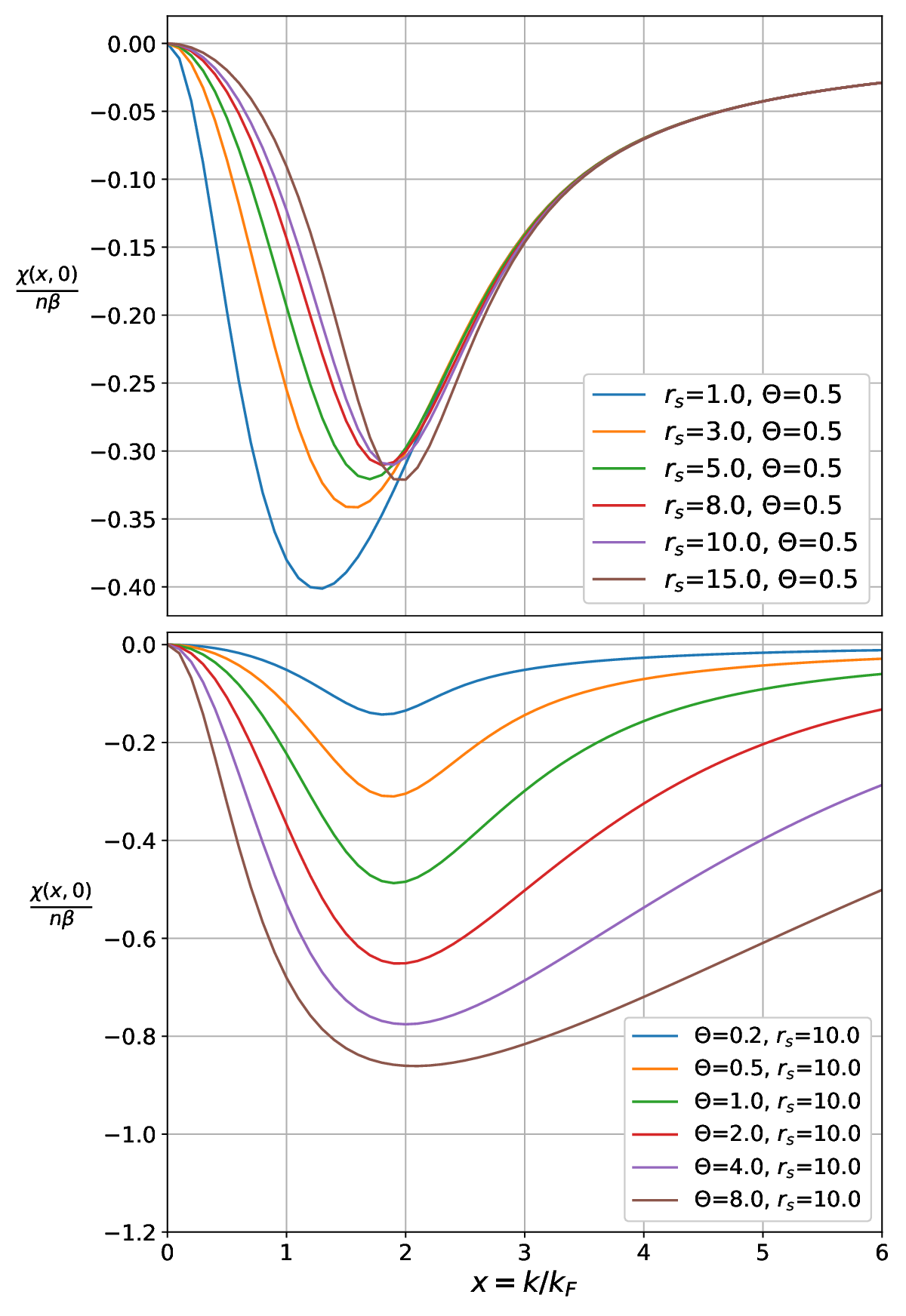}
	\caption{The dependence of the static density response of the paramagnetic warm dense uniform electron fluid on the thermodynamic variables $(r_{\mathrm{s}},\Theta)$, as predicted by the finite temperature VS scheme. (Top) Results for constant $\Theta=0.5$ and for varying $r_{\mathrm{s}}=\{1,3,5,8,10,15\}$. (Bottom) Results for constant $r_{\mathrm{s}}=10$ and for varying $\Theta=\{0.2,0.5,1.0,2.0,4.0,8.0\}$.}\label{SDR_VSparametric}
\end{figure}

The dependence of the VS-generated SDRs on the UEF thermodynamic variables $(r_{\mathrm{s}},\Theta)$ is explored in Fig.\ref{SDR_VSparametric}. At constant $\Theta$ and for increasing $r_{\mathrm{s}}$; the short wavenumber behavior of the SDR is characterized by a very strong dependence as expected from the normalized unit version $\lim_{x\to0}\chi(\boldsymbol{x})/(n\beta)=-(3\pi/8\lambda)(\Theta/r_{\mathrm{s}})x^2$ of the perfect screening asymptote. On the other hand, all the SDRs exhibit the same large wavenumber behavior given the $r_{\mathrm{s}}-$independence of the ideal SDR with the $r_{\mathrm{s}}=15$ curve attaining its limiting behavior at the highest wavenumber. The position of the omnipresent SDR minimum gets gradually shifted to larger wavenumbers, as the quantum coupling parameter increases. The magnitude of the SDR minimum is monotonically decreasing with $r_{\mathrm{s}}$ up to the emergence of the maximum of the SSF. At even higher coupling, the magnitude of the SDR minimum begins to increase with $r_{\mathrm{s}}$\,\cite{newQMCd,sdrref6}. At constant $r_{\mathrm{s}}$, there is a strong dependence of both asymptotic limits on $\Theta$. It is emphasized that the position of the SDR minimum weakly depends on the degeneracy parameter, whereas the SDR minimum becomes broader as $\Theta$ increases.

\subsection{Pair correlation function}\label{subsec:PCFresults}

\begin{figure*}
	\centering
	\includegraphics[width=7.0in]{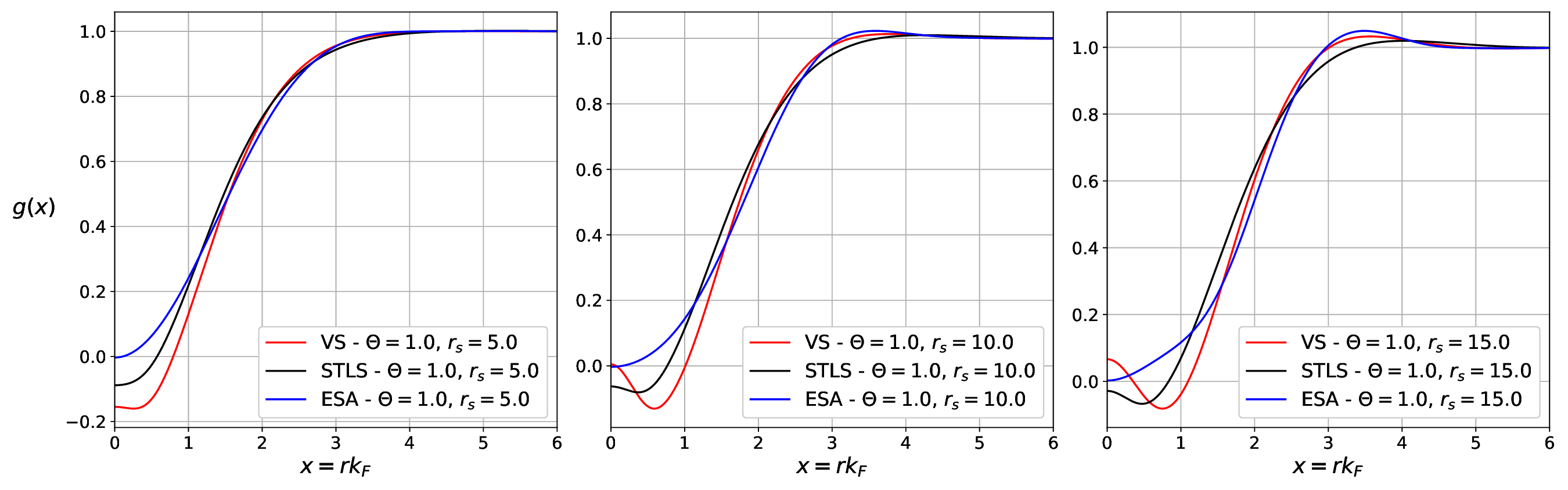}
	\caption{The pair correlation function of the paramagnetic uniform electron fluid in the warm dense matter regime, as predicted by the VS scheme (red), the STLS scheme (black) and the quasi-exact ESA scheme (blue). Results for $\Theta=1.0$ and for $r_{\mathrm{s}}=5$ (left), $r_{\mathrm{s}}=10$ (center), $r_{\mathrm{s}}=15$ (right).}\label{PCF_VSvsSTLS}
\end{figure*}

\begin{figure}
	\centering
	\includegraphics[width=3.50in]{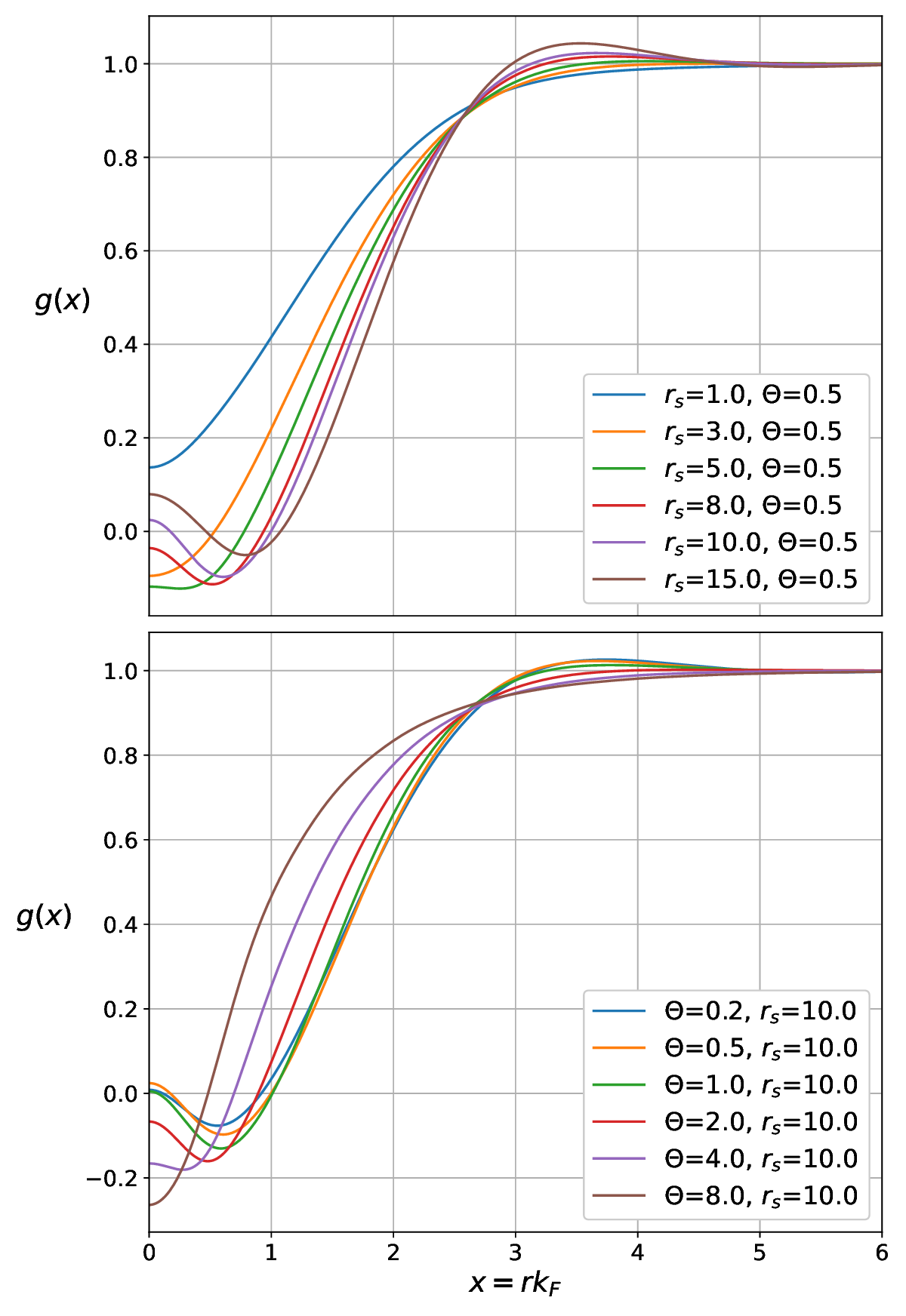}
	\caption{Dependence of the pair correlation function of the paramagnetic warm dense uniform electron fluid on the thermodynamic variables $(r_{\mathrm{s}},\Theta)$, as predicted by the finite temperature VS scheme. (Top) Results for constant $\Theta=0.5$ and for varying $r_{\mathrm{s}}=\{1,3,5,8,10,15\}$. (Bottom) Results for constant $r_{\mathrm{s}}=10$ and for varying $\Theta=\{0.2,0.5,1.0,2.0,4.0,8.0\}$.}\label{PCF_VSparametric}
\end{figure}

\noindent The PCF is obtained from the SSF by the inverse Fourier transform of the fundamental microscopic expression $S(\boldsymbol{k})=1+nH(\boldsymbol{k})$\,\cite{IETliqu}. For isotropic systems such as the UEF, use of spherical coordinates and normalized units leads to
\begin{equation}
g(x)=1+\frac{3}{2x}\int_0^{\infty}[S(y)-1]y\sin{(xy)}dy\,.\label{PCF_from_SSF}
\end{equation}
It is emphasized that specialized quadrature algorithms are necessary for accurate numerical evaluation of this integral at short distances owing to the rapid oscillatory nature of the $\sin{(xy)}$ factor of the integrand\,\cite{pcfref1,pcfref2,pcfref3}. It is also pointed out that the truncation of the upper integration limit should be handled with care; the SSF reaches its unity asymptotic limit around $y\sim5$ and thus it is possible that noise of alternating sign is solely integrated at large wavenumbers for which the SSF fluctuates around unity with magnitudes that are lower than the accuracy of the algorithm employed for the solution of the dielectric scheme of interest.

A comparison of the PCF predictions of the STLS, VS and ESA schemes is featured in Fig.\ref{PCF_VSvsSTLS}. It is preferable to distinguish between the near-contact short distance region (which is primarily shaped by the competition between quantum effects that allow the overlap of electrons of
opposite spin\,\cite{pcfref4} and strong interaction effects that favor the formation of correlation voids\,\cite{pcfref5,pcfref6}) and the intermediate-long distance region that consists of the first coordination shell (which is primarily shaped by the relative strength of the correlations\,\cite{IETliqu,normal1}) together with the asymptotic decay range (where the PCF reaches its long distance limit of unity\,\cite{IETliqu}). \textbf{(i)} Let us first consider the combined intermediate-long distance region $rk_{\mathrm{F}}\gtrsim1$. At high densities, the STLS-generated and VS-generated PCFs have comparable accuracy and exhibit small deviations from the quasi-exact ESA PCF that are spread over all distances $rk_{\mathrm{F}}\lesssim4$. At stronger coupling, the ESA PCF has a shallow maximum whose position is accurately predicted by the VS scheme, unlike its magnitude which is underestimated by the VS scheme. On the other hand, the STLS scheme leads to an even shallower PCF maximum that is displaced towards larger distances. Thus, STLS-generated PCFs may attain their asymptotic limit at larger distances than VS- and ESA-generated PCFs. \textbf{(ii)} Let us next proceed with the discussion of the short distance region $rk_{\mathrm{F}}\lesssim1$. The STLS and the VS schemes lead to an un-physical negative PCF region near the origin ($r=0$). This is a well-known common pathological feature of all non-empirical schemes of the dielectric formalism that originates from the approximate treatment of quantum effects\,\cite{DornRev,pcfref8,pcfref9}. It is particularly prominent in semi-classical schemes, but it is also present in pure quantum schemes as well as in quantized schemes based on integral equation theories\,\cite{IETChem,qIETLet}. On the other hand, mapping approaches that are based integral equation theories for an effective pair interaction potential within a specific classical-quantum state correspondence rule by default lead to a non-negative PCF\,\cite{pcfref8,pcfref9}. As expected based on their semi-classical classification, both the VS and STLS schemes are characterized by extended negative PCF regions. For most state points, the absolute value of the integrated negative region of the VS-generated PCF is larger than the respective absolute value of the STLS-generated PCF. \textbf{(iii)} Let us also focus on the contact ($r=0$). The larger integrated negative PCF region of the VS scheme does not necessarily translate to a more negative on-top PCF, $g(r=0)$, owing to the existence of a still unphysical negative global minimum of the PCF. The latter is observed for both the VS and the STLS schemes from moderate coupling. More specifically, even though the VS minimum is deeper, the larger VS slope at the contact-side of the minimum yields less negative (or even positive) on-top PCF values compared to the STLS scheme. \textbf{(iv)} For completeness, it is noted that the semi-empirical nature of the ESA scheme allows it to be an exception to the negative PCF pathology of dielectric schemes. It incorporates accurate QMC-based on-top PCF values and, for nearly all state points within its applicability range, yields a PCF that is monotonic prior to the first coordination shell maximum implying that ESA-generated PCFs are always positive.

The dependence of the VS-generated PCFs on the UEF thermodynamic variables $(r_{\mathrm{s}},\Theta)$ is explored in Fig.\ref{PCF_VSparametric}. Inside the first coordination shell, the $r_{\mathrm{s}}-$dependence and the $\Theta-$dependence of the PCF can be fully understood on the basis of the earlier SSF discussion at the last paragraph of Sec.\ref{subsec:SSFresults}. Concerning short distances; the PCF is always positive and monotonic at high electron densities, the onset (but not necessarily the extent) of the negative PCF region is observed to monotonically increase as $r_{\mathrm{s}}$ rises and monotonically increase as $\Theta$ decreases up to $0.5\lesssim\Theta\lesssim1$, the negative PCF minimum is not present for low effective coupling parameters (high $\Theta$ and/or low $r_{\mathrm{s}}$) and the on-top PCF values are positive either at high densities (non-negative PCF) or for large effective coupling parameters (negative PCF region with a minimum).

\subsection{Static local field correction}\label{subsec:SLFCresults}

\begin{figure*}
	\centering
	\includegraphics[width=7.0in]{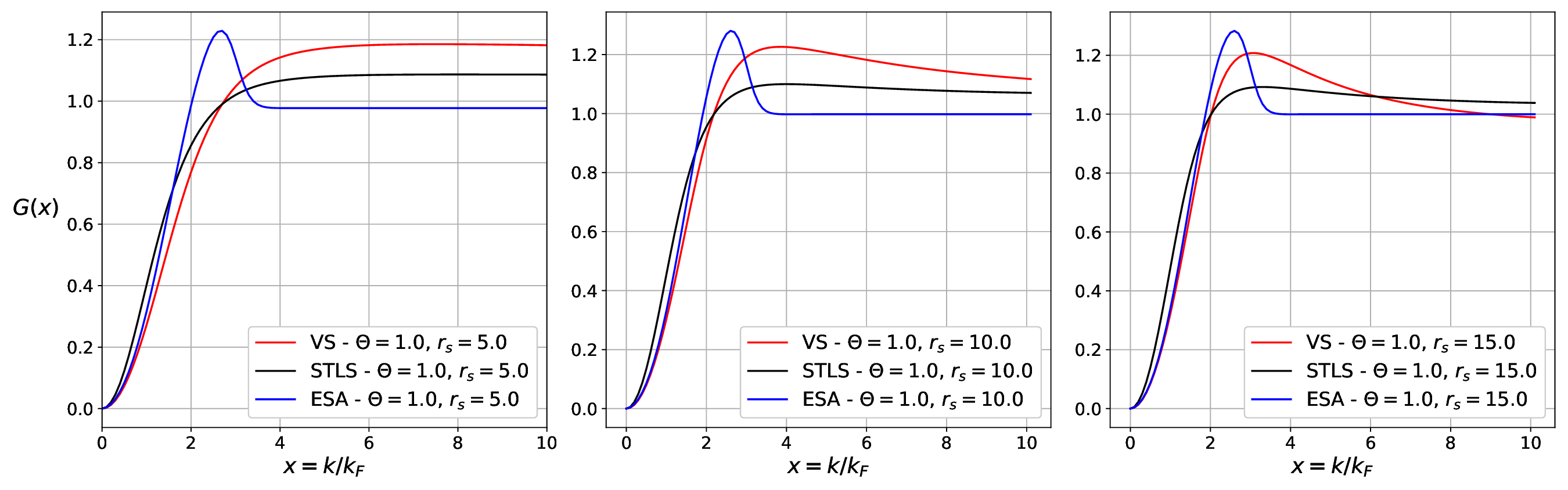}
	\caption{Static local field correction of the paramagnetic uniform electron fluid in the warm dense matter regime, as predicted by the VS scheme (red), the STLS scheme (black) and the ESA scheme (blue). Results for $\Theta=1.0$ and for $r_{\mathrm{s}}=5$ (left), $r_{\mathrm{s}}=10$ (center), $r_{\mathrm{s}}=15$ (right).}\label{SLFC_VSvsSTLS}
\end{figure*}

\begin{figure}
	\centering
	\includegraphics[width=3.50in]{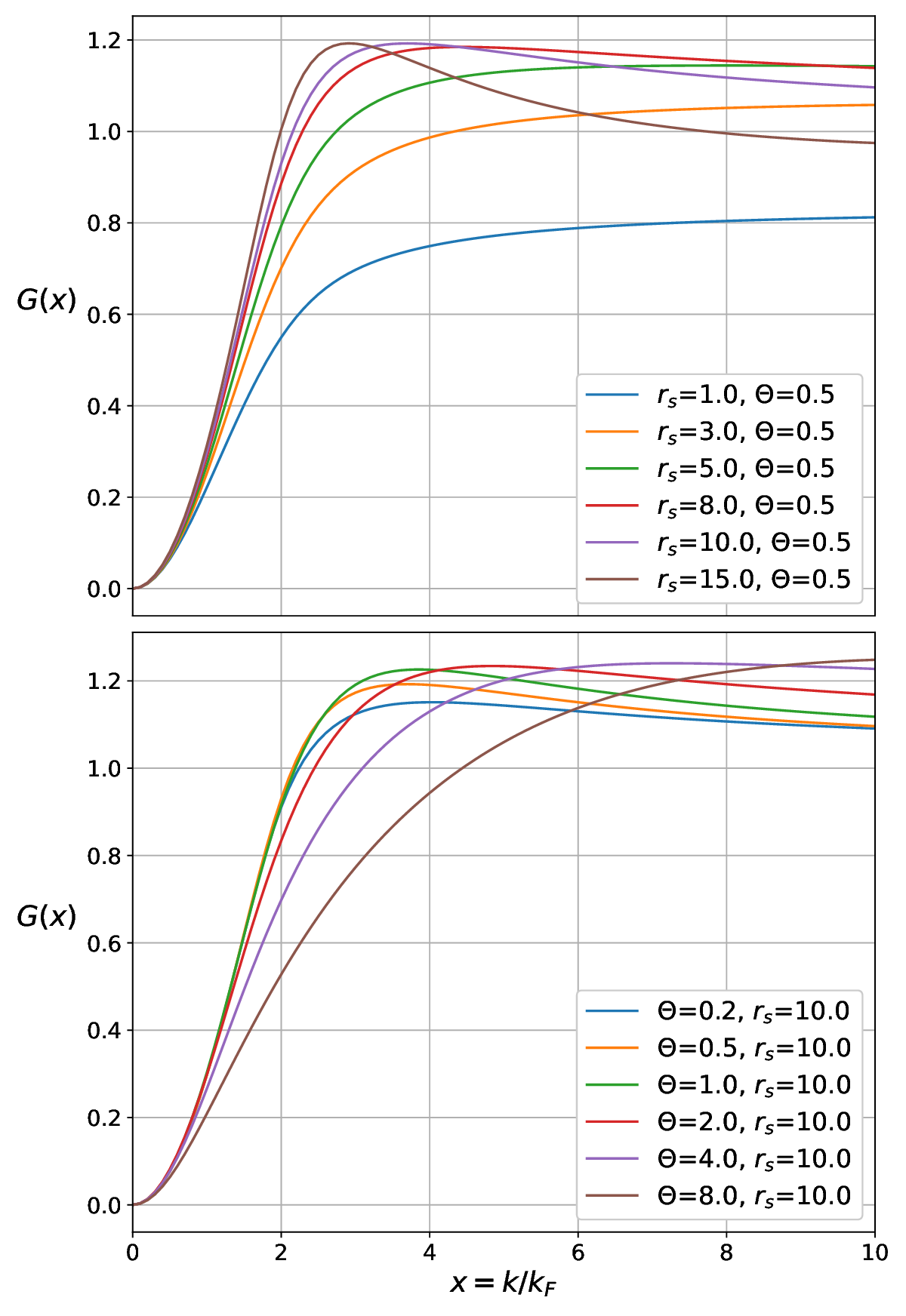}
	\caption{Dependence of the static local field correction of the paramagnetic warm dense uniform electron fluid on the thermodynamic variables $(r_{\mathrm{s}},\Theta)$, as predicted by the finite temperature VS scheme. (Top) Results for constant $\Theta=0.5$ and for varying $r_{\mathrm{s}}=\{1,3,5,8,10,15\}$. (Bottom) Results for constant $r_{\mathrm{s}}=10$ and for varying $\Theta=\{0.2,0.5,1.0,2.0,4.0,8.0\}$.}\label{SLFC_VSparametric}
\end{figure}

\noindent Provided that the local field correction is static, its large wavenumber limit is algebraically connected to the on-top PCF. After the substitution of $G(\boldsymbol{k},\omega)\equiv{G}(\boldsymbol{k})$ in the constitutive relation for the density response function Eq.(\ref{densityresponseDLFC}), the combination with the infinite Matsubara summation of Eq.(\ref{Matsubaraseries}) and the utilization of the fundamental PCF-SSF connection, an intricate asymptotic analysis yields the general result\,\cite{IchiRep,TanIchi}
\begin{equation}
\left.\frac{\partial{g}(r)}{\partial{r}}\right|_{r=0}=\frac{1}{a_{\mathrm{B}}}\left[1-\lim_{k\to\infty}G(\boldsymbol{k})\right]\,.\label{IchiCusp}
\end{equation}
Simultaneously, the cusp condition is exactly valid which states that the derivative of the PCF logarithm at the origin equals the inverse Bohr radius\,\cite{ESApap4,slfcre1,slfcre2}, i.e.,
\begin{equation}
\left.\frac{\partial{g}(r)}{\partial{r}}\right|_{r=0}=\frac{g(0)}{a_{\mathrm{B}}}\,.\label{KimbCusp}
\end{equation}
Combining the above, one obtains the exact cusp relation that reads as\,\cite{IchiRep,TanIchi}
\begin{equation}
\lim_{k\to\infty}G(\boldsymbol{k})=1-g(0)\,.\label{FullCusp}
\end{equation}
This can be considered as a self-consistency condition\,\cite{IchiRep,TanIchi} since it solely originates from the first two exact building blocks of the dielectric formalism, see Eqs.(\ref{densityresponseDLFC},\ref{Matsubaraseries}), and needs to be independently satisfied by the third approximate building block, see the closure functional of Eq.(\ref{functionalclosure}) without the frequency dependence.

It can be easily shown that the STLS scheme satisfies the cusp self-consistency relation. The large wavenumber limit $x\to\infty$ of the bracketed factor of the STLS SLFC integrand, see Eq.(\ref{STLSSLFCftcorr}), is found to be equal to $2$, after a Taylor expansion with respect to $y/x\to0$. This leads to
\begin{align*}
G_{\mathrm{STLS}}(x)&=-\frac{3}{2}\int_0^{\infty}dyy^2\left[S(y)-1\right]\,,
\end{align*}
where the state point dependence has been suppressed. The short distance limit $x\to0$ of the fundamental PCF-SSF connection, see Eq.(\ref{PCF_from_SSF}), since $\sin{(xy)}\simeq{xy}$, yields
\begin{align*}
g(0)=1+\frac{3}{2}\int_0^{\infty}y^2[S(y)-1]dy\,.
\end{align*}
Thus, combining the above, one indeed obtains that
\begin{equation}
\lim_{x\to\infty}G_{\mathrm{STLS}}(x)=1-g(0)\,.\label{FullCuspSTLS}
\end{equation}
It is rather straightforward to prove that the VS scheme cannot satisfy the cusp self-consistency relation. In view of the differential connection between the VS and STLS SLFCs, see Eq.(\ref{VSSLFCgen}) or Eq.(\ref{VSSLFCftcorr}), and courtesy of the asymptotic $\left.x(\partial{G_{\mathrm{STLS}}(x)}/{\partial{x}})\right|_{x\to\infty}=0$, we directly obtain\,\cite{ESApap3}
\begin{equation}
\lim_{x\to\infty}G_{\mathrm{VS}}(x)=1-g(0)+\frac{2}{3}\alpha\Theta\frac{\partial{g}(0)}{\partial\Theta}+\frac{1}{3}\alpha{r}_{\mathrm{s}}\frac{\partial{g}(0)}{\partial{r}_{\mathrm{s}}}\,.\label{FullCuspVS}
\end{equation}
From general physical arguments that are based on the non-interacting limit and the strongly interacting limit, it can be deduced that the thermodynamic derivatives of the on-top PCF cannot be identically zero. In addition, based on the extended discussion featured at the last paragraph of Sec.\ref{subsec:PCFresults}, it can be safely concluded that the thermodynamic derivatives of the on-top PCF do not provide negligible contributions to Eq.(\ref{FullCuspVS}). Therefore, the VS scheme does not even roughly comply with the cusp self-consistency relation. In a sense, enforcement of small wavenumber consistency (satisfaction of isothermal compressibility sum rule) has the undesired side-effect of large wavenumber inconsistency (violation of cusp condition).

\begin{figure*}
	\centering
	\includegraphics[width=6.0in]{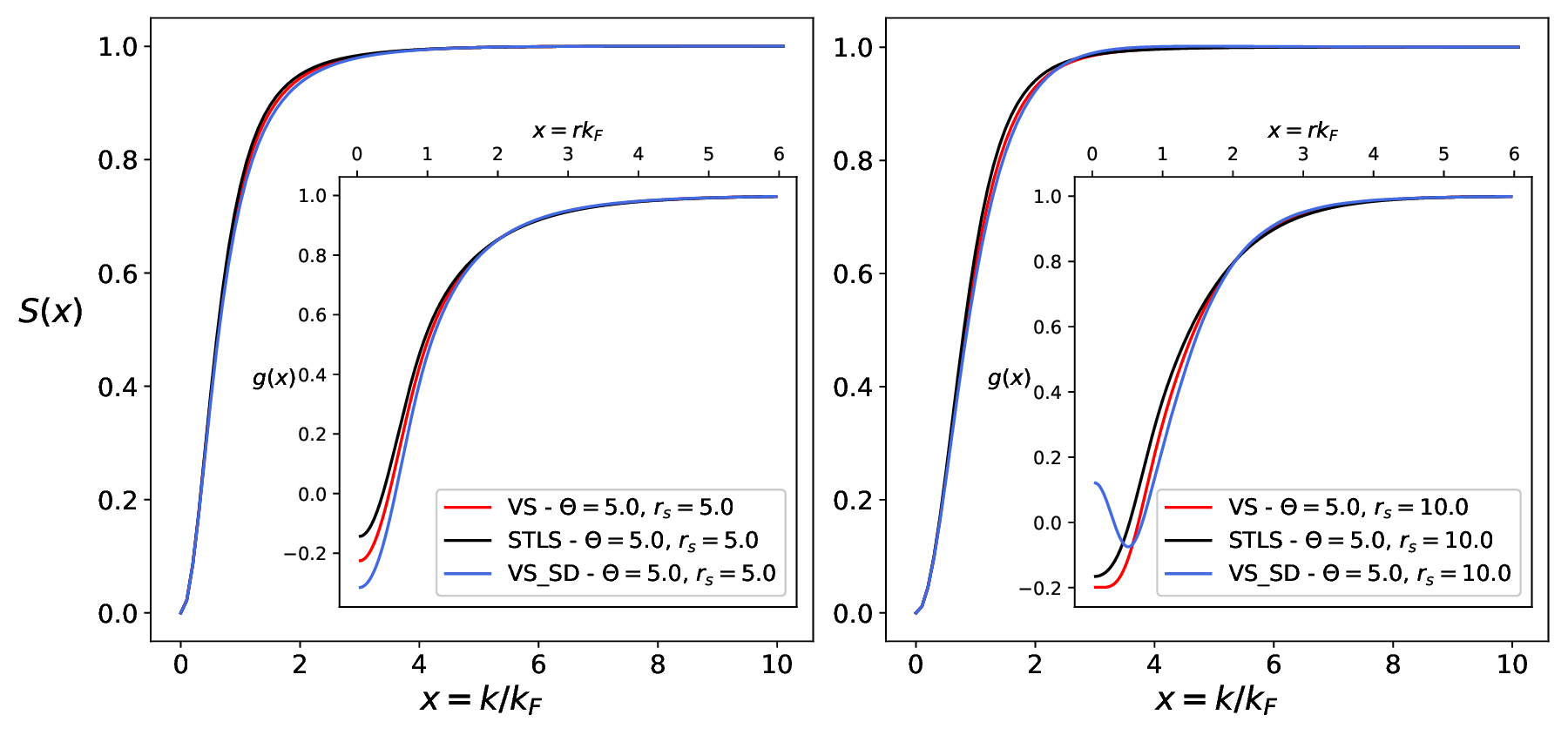}
	\caption{Static structure factor (main) and pair correlation function (inset) of the paramagnetic uniform electron fluid in the warm dense matter regime, as predicted by the VS-SD scheme (blue), VS scheme (red) and STLS scheme (black). Results for $\Theta=5.0$ and $r_{\mathrm{s}}=5$ (left), $r_{\mathrm{s}}=10$ (right).}\label{SSF_VSvsVSSD}
\end{figure*}

A comparison of the SLFC predictions of the STLS, VS and ESA schemes is featured in Fig.\ref{SLFC_VSvsSTLS}. It is pointed out that the ESA scheme nearly exactly satisfies the cusp self-consistency relation and the CSR expression by construction\,\cite{ESApap2}. To prevent possible misconceptions, it is also emphasized that the ESA scheme features a frequency-averaged LFC that does not correspond to the exact static limit of the dynamic LFC as that would emerge from QMC simulations\,\cite{ESApap1,ESApap2}. \textbf{(i)} Concerning the SLFC long wavelength limit, as expected from its direct connection with the CSR, VS-generated SLFCs are more accurate than STLS-generated SLFCs regardless of the UEF state point. Furthermore, as the quantum coupling parameter increases, the SLFC result exhibits less deviations from the accurate ESA result. This is consistent with the improved agreement of the VS exchange correlation free energy with the GDSMFB parametrization as $r_{\mathrm{s}}$ increases. \textbf{(ii)} Concerning the large wavenumber limit of the SLFC, as expected from its connection with the cusp condition, the VS-generated SLFCs are less accurate than the STLS-generated SLFCs regardless of the UEF state point of interest. Regarding the STLS scheme and the ESA scheme, the cusp self-consistency relation also provides an independent check of the numerical accuracy of the respective algorithms; the STLS SLFC large wavenumber limit is indeed larger than unity in accordance with the negative on-top PCF, whereas the ESA SLFC large wavenumber limit is indeed smaller (or nearly equal to) than unity in accordance with the non-negative on-top PCF. \textbf{(iii)} Concerning the intermediate wavenumber region, the deviations from the ESA SLFC are always rather large, but it can be stated that the STLS SLFC is more accurate at high densities and that the VS SLFC becomes more accurate as the coupling increases. The STLS SLFC is characterized by a shallow maximum only near the boundary between the WDM and the strongly coupled regimes, while the VS SLFC only lacks a maximum at high densities. Yet, even at the largest coupling parameters probed, the well-formed VS SLFC maximum could not match the magnitude, position and width of the ESA SLFC maximum. \textbf{(iv)} In contrast to the sharp ESA SLFC transition between the intermediate wavenumber region and the asymptotic value, in case a SLFC maximum is present, the respective VS and STLS SLFC transitions are slow and could even extend over several Fermi wavenumbers.

The dependence of VS-generated SLFCs on the UEF thermodynamic variables $(r_{\mathrm{s}},\Theta)$ is explored in Fig.\ref{SLFC_VSparametric}. The strong dependence of the large wavenumber limit on $(r_{\mathrm{s}},\Theta)$ reflects the strong dependence of the on-top PCF on $(r_{\mathrm{s}},\Theta)$, in view of Eq.(\ref{FullCuspVS}). The aforementioned correlation between the presence of a SLFC maximum and the prolonged asymptotic transition is also apparent. Finally, the weak dependence of the long wavelength limit $(r_{\mathrm{s}},\Theta)$, except from high $\Theta$ and/or low $r_{\mathrm{s}}$, reflects the relatively weak dependence of the exchange-correlation free energy on $(r_{\mathrm{s}},\Theta)$, in view of Eq.(\ref{VSCSRgen}).

\subsection{Comparison with the VS-SD scheme}\label{subsec:VSSDresults}

\noindent Owing to the erroneous chain rule application in the density derivative, the Sjostrom and Dufty formulation of the finite temperature VS scheme (VS-SD scheme)\,\cite{VSsche2} is considerably simpler than the present corrected formulation of the finite temperature VS scheme. First and foremost, when utilizing second order finite difference schemes to approximate the thermodynamic derivatives, the minimum VS-SD computational stencil features three state points, while the minimum corrected VS computational stencil features nine state points, compare Eqs.(\ref{VSSLFCftcorr},\ref{VSSLFCftSD}). As a consequence, the inner loop of the VS-SD algorithm needs to be simultaneously run for three state points, while the inner loop of the VS algorithm needs to be simultaneously run for nine state points. Thus, the VS-SD computational cost is at least three times smaller. In addition, the self-consistency parameter equation of the VS-SD scheme involves considerably less thermodynamic derivatives than the self-consistency parameter equation of the VS scheme, compare Eqs.(\ref{selfconsistencyVS},\ref{selfconsistencyVSSD}). This directly suggests less complexity and indirectly implies less computational cost due to the faster convergence of the outer loop.

In this subsection, we compare the predictions of the (corrected) VS scheme with the predictions of the VS-SD scheme. Such a comparison allows us to confirm that the corrected treatment has an meaningful impact on the structural properties of the UEF and is not merely an added complexity. Such a comparison also allows us to identify phase diagram regions wherein the VS scheme is indistinguishable with the VS-SD scheme and thus the simpler version can be employed without introducing errors. After a simple inspection of the respective SLFC closures and respective self-consistency parameter equations, it can be deduced that the two schemes will have maximum differences at high $\Theta$ and/or small $r_{\mathrm{s}}$ and will nearly overlap at low $\Theta$ and/or large $r_{\mathrm{s}}$. On the other hand, the structural properties are not very sensitive to typical values of the quantum coupling parameter that are relevant for WDM applications ($r_{\mathrm{s}}\sim5$), see Figs.\ref{SSF_VSparametric},\ref{SDR_VSparametric}.

\begin{figure*}
	\centering
	\includegraphics[width=6.0in]{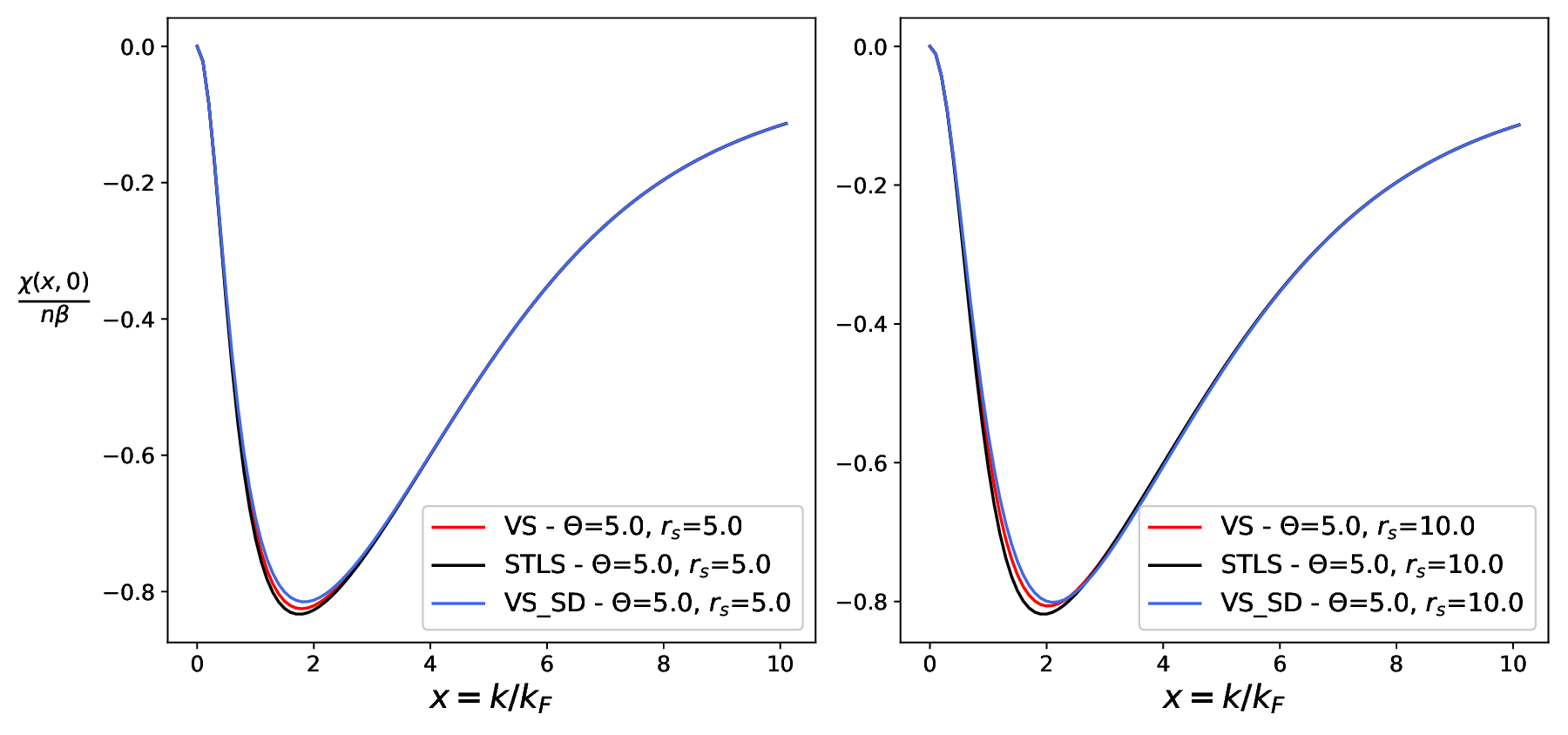}
	\caption{Static density response of the paramagnetic uniform electron fluid in the warm dense matter regime, as predicted by the VS-SD scheme (blue), VS scheme (red) and STLS scheme (black). Results for $\Theta=5.0$ and $r_{\mathrm{s}}=5$ (left), $r_{\mathrm{s}}=10$ (right).}\label{SDR_VSvsVSSD}
\end{figure*}

\begin{figure*}
	\centering
	\includegraphics[width=6.0in]{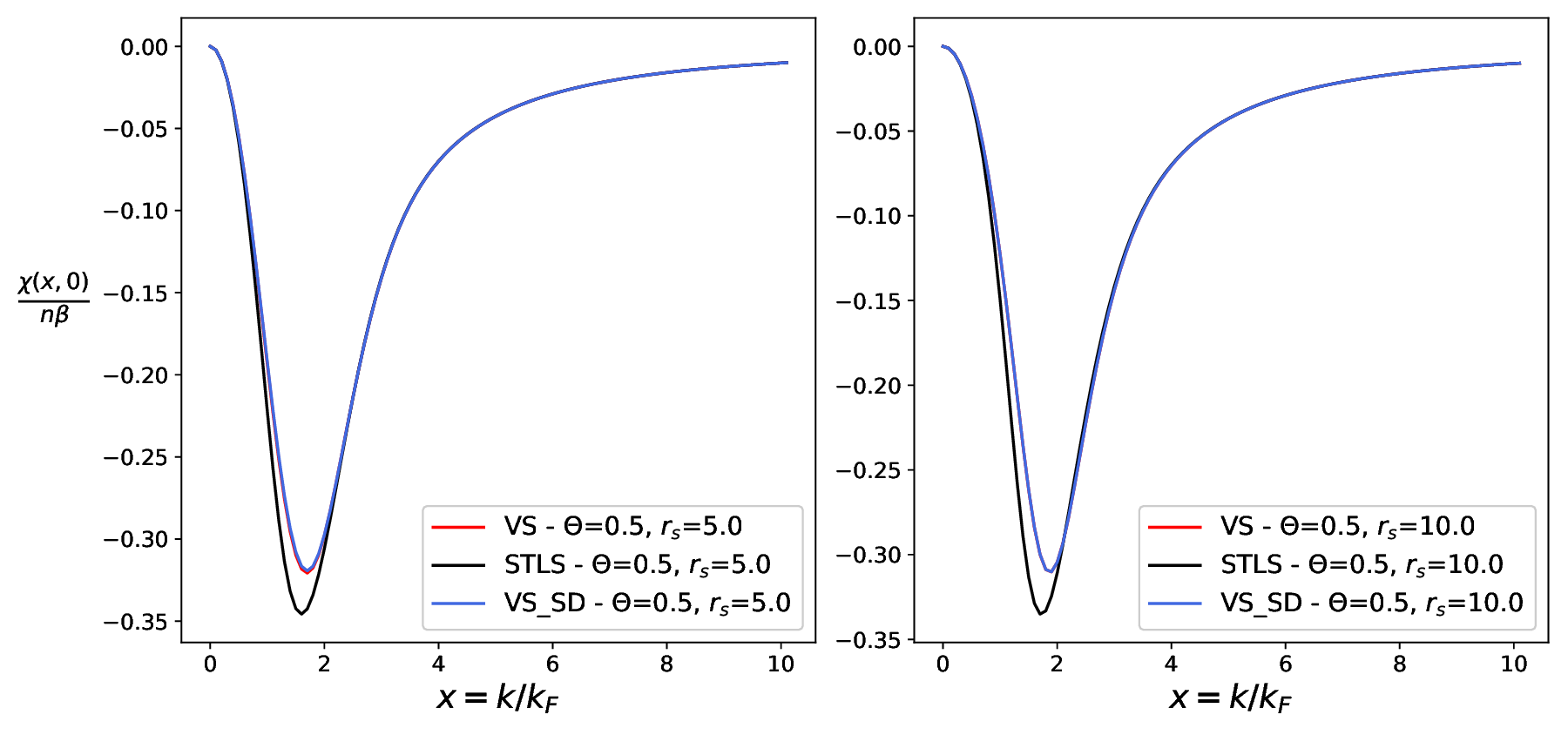}
	\caption{Static density response of the paramagnetic uniform electron fluid in the warm dense matter regime, as predicted by the VS-SD scheme (blue), VS scheme (red) and STLS scheme (black). Results for $\Theta=0.5$ and $r_{\mathrm{s}}=5$ (left), $r_{\mathrm{s}}=10$ (right).}\label{SDR_VSsameVSSD}
\end{figure*}

A comparison of the SSF and PCF predictions of the STLS, VS and VS-SD schemes for $\Theta=5$ and $r_{\mathrm{s}}=5,10$ is featured in Fig.\ref{SSF_VSvsVSSD}. In line with our expectations, the VS-SD scheme leads to unique predictions for both structural properties. In particular, as discerned in the main figures, the VS-generated SSF consistently lies in-between the STLS (upper bound) and VS-SD (lower bound) predictions. Furthermore, as illustrated in the insets, there are substantial PCF differences between the VS and the VS-SD schemes that are mainly confined to short distances. Finally, the inset of the right panel even reveals that the unphysical negative PCF region near the origin can be monotonic for the VS scheme and non-monotonic for the VS-SD scheme.

A comparison of the SDR predictions of the STLS, the VS and the VS-SD schemes for $\Theta=5$ and $r_{\mathrm{s}}=5,10$ is featured in Fig.\ref{SDR_VSvsVSSD}. Unsurprisingly, the VS-SD scheme yields distinct predictions for the static limit of the linear density response function. For both coupling parameters, the VS-generated SDR lies in-between the STLS (lower bound) and the VS-SD (upper bound) predictions. In general, this is a pattern that persists within the WDM regime. Therefore, it can be stated that the correct finite temperature formulation of the VS scheme leads to predictions that are closer to the STLS scheme compared to the predictions of the VS-SD scheme. Another comparison of the SDR predictions of the STLS, VS and VS-SD schemes now for $\Theta=0.5$ and $r_{\mathrm{s}}=5,10$ is featured in Fig.\ref{SDR_VSsameVSSD}. The VS and VS-SD results for the SDR are truly indistinguishable, overlapping in the entire wavenumber range. The same applies for the PCF, SSF and SLFC.

\section{Summary and future work}

\subsection{Summary}\label{subsec:summary}

\noindent In this work, the finite temperature formulation of the Vashishta--Singwi dielectric formalism scheme for the uniform electron fluid was reported correcting for a thermodynamic derivative error that was present in the earlier formulation of Sjostrom and Dufty. This is a truly self-consistent version, where the density expansion parameter $\alpha$ is not locked to an approximate state independent value but is determined by enforcing the compressibility sum rule at each thermodynamic point. An efficient computational scheme was devised for the solution of the VS scheme that was presented in full detail. It features an inner loop that simultaneously solves the scheme for nine state points, which form a minimum extent computational stencil when all finite difference approximations of the thermodynamic derivatives are of the second order. It also features an outer loop that determines the self-consistency parameter $\alpha$ through a non-linear equation of the form $\alpha=f_1[\widetilde{f}_{\mathrm{xc}}(\alpha)]/f_2[\widetilde{u}_{\mathrm{int}}(\alpha)]$, where $f_1$ contains all possible first- and second-order thermodynamic derivatives of $\widetilde{f}_{\mathrm{xc}}$ and $f_2$ contains all possible first-order thermodynamic derivatives of $\widetilde{u}_{\mathrm{int}}$, which is solved with the secant method. The VS algorithm is implemented in a home-made hybrid code, where C++ is used for the backend and python is used for the frontend.

The finite temperature VS scheme was numerically solved for $2250$ state points of the paramagnetic uniform electron fluid that cover the entire warm dense matter regime. First, this extended parametric study facilitated the extraction of an accurate closed-form expression for the self-consistency parameter $\alpha(r_{\mathrm{s}},\Theta)$ whose availability removes the necessity for an outer loop and permits the use of STLS-like algorithms for the solution of the VS scheme. Furthermore, this systematic investigation provided sufficient data for the determination of an accurate closed-form expression for the exchange correlation free energy following the lines of state-of-the-art parametrizations, which adequately represents the thermodynamic predictions of the VS scheme. More important, this allowed a comprehensive thermodynamic and structural comparison with the respective predictions of the ubiquitous STLS scheme as well as the near-exact thermodynamic results of the QMC-based GDSMFB equation of state and the near-exact structural results of the QMC-based effective static approximation.

The primary conclusions drawn from the aforementioned comparison are summarized in the following: \textbf{(i)} In the entire WDM regime, the thermodynamic predictions of the STLS scheme are more accurate than the thermodynamic predictions of the VS scheme benefitting from a very favorable cancellation of errors in the static structure factor integral for the interaction energy. \textbf{(ii)} In the high density range of the WDM regime, $r_{\mathrm{s}}\lesssim5$ at the Fermi temperature, the STLS static structure factor is consistently more accurate than the VS static structure factor. The situation reverses at the heart of the WDM regime and towards its boundary with the strongly coupled regime, especially at the long wavelength limit and at the vicinity of the static structure factor maximum. This observation can be explained by the fact that the VS scheme indirectly includes ternary correlations through the density derivative of the pair correlation function. \textbf{(iii)} As far as the static density response is concerned, both long wavelength and large wavenumber asymptotic limits are automatically satisfied by both schemes, but the STLS scheme yields more accurate predictions for the magnitude of the omnipresent minimum (which progressively worsen as the coupling increases), while the VS scheme yields more accurate predictions for the position of the omnipresent minimum. \textbf{(iv)} The structural superiority of the VS scheme at the strongly coupled region of the WDM regime and the structural superiority of the STLS scheme at the high density region of the WDM regime are also confirmed from the pair correlation function and the static local field correction. \textbf{(v)} An unphysical negative region in the pair correlation function near contact is present for both schemes. This does not necessarily translate to negative on-top values, because, at strong coupling, the pair correlation function can exhibit a negative minimum. For most state points, the absolute value of the integrated negative region is larger in the VS scheme than in the STLS scheme. This short distance drawback of the VS scheme most probably originates from the violation of the cusp condition, which is exactly satisfied by the STLS scheme.

Finally, our correct formulation of the VS scheme collapses to the earlier Sjostrom and Dufty formulation of the VS scheme at low values of the degeneracy parameter and/or large values of the quantum coupling parameter. This was expected by the nature of the thermodynamic derivative error of the Sjostrom and Dufty analysis and has been confirmed numerically. Thus, there is still some utility in the less computationally costly and easier to numerically implement Sjostrom and Dufty formulation.

\subsection{Future work}\label{subsec:future}

\noindent One of the main drawbacks of the finite temperature and ground state versions of the VS scheme concerns its semiclassical nature, i.e., the treatment of quantum mechanical effects at the level of the random phase approximation. However, the standard VS closure can be employed to truncate the first member of the quantum BBGKY hierarchy, which would allow the incorporation of quantum effects beyond the Vlasov-Lindhard ideal response substitution that would then be represented by a dynamic LFC (DLFC). We shall refer to this dielectric scheme as the quantum VS (qVS). The standard perturbative analysis of the quantum kinetic equation within the Wigner representation leads to the qVS DLFC that reads as
\begin{align}
&G_{\mathrm{qVS}}(\boldsymbol{k},\omega)=\left(1+{\alpha}n\displaystyle\frac{\partial}{\partial{n}}\right)G_{\mathrm{qSTLS}}(\boldsymbol{k},\omega)\,,\nonumber\\
&G_{\mathrm{qSTLS}}(\boldsymbol{k},\omega)=-\displaystyle\frac{1}{n}\int\frac{d^3k^{\prime}}{(2\pi)^3}\frac{k^2}{{k^{\prime}}^2}\frac{\chi_0(\boldsymbol{k},\boldsymbol{k}^{\prime},\omega)}{\chi_0(\boldsymbol{k},\omega)}\left[S(\boldsymbol{k}-\boldsymbol{k}^{\prime})-1\right]\,,\nonumber\\
&\chi_0(\boldsymbol{k},\boldsymbol{k}^{\prime},\omega)=-\frac{2}{\hbar}\int\frac{d^3q}{(2\pi)^3}\frac{f_0\left(\boldsymbol{q}+\frac{1}{2}\boldsymbol{k}^{\prime}\right)-f_0\left(\boldsymbol{q}-\frac{1}{2}\boldsymbol{k}^{\prime}\right)}{\omega-\frac{\hbar}{m}\boldsymbol{k}\cdot\boldsymbol{q}+\imath0}\,.\nonumber
\end{align}
In these expressions; $G_{\mathrm{qSTLS}}(\boldsymbol{k},\omega)$ denotes the DLFC of the quantum STLS (qSTLS) scheme, which emerges by truncating the first member of the quantum BBGKY hierarchy with the standard STLS closure\,\cite{qSTLSge,qSTLSgr,qSTLSFT}, while $\chi_0(\boldsymbol{k},\boldsymbol{k}^{\prime},\omega)$ denotes the three-argument ideal density response function which collapses to the ideal (Lindhard) density response function when $\boldsymbol{k}^{\prime}=\boldsymbol{k}$, $\chi_0(\boldsymbol{k},\boldsymbol{k},\omega)=\chi_0(\boldsymbol{k},\omega)$\,\cite{qSTLSge,qSTLSgr,qSTLSFT}. Note that the correspondence between the qVS DLFC, qSTLS DLFC is exactly the same with the correspondence between the VS SLFC, STLS SLFC, compare the above with Eqs.(\ref{VSSLFCgen},\ref{STLSSLFCgen}). It is also important to emphasize that the qVS DLFC can be directly obtained from the VS SLFC after the substitution $(\boldsymbol{k}\cdot\boldsymbol{k}^{\prime})/{k^2}\to\chi_0(\boldsymbol{k},\boldsymbol{k}^{\prime},\omega)/\chi_{0}(\boldsymbol{k},\omega)$ within the wavenumber integrand; this substitution rule was first observed in Ref.\cite{qIETLet} where it was utilized to rapidly quantize semi-classical non-perturbative schemes\,\cite{HNCSTLS,IETChem,IETLett}. To our knowledge, the ground state version of the qVS scheme has only been considered with a locked parameter $\alpha=1/2$ in the literature\,\cite{qSTLSgr,qSTLSex}, which implies that the CSR rule is slightly violated. In the fully self-consistent case, the transition from the present VS scheme to the envisaged qVS scheme will be accompanied by a (manageable) blow-up in the computational cost of the inner loop. In particular, the dynamic nature of the LFC implies that the qVS closure equation depends on the Matsubara frequencies, while the presence of the three-argument ideal density response implies that the qVS closure equation is now described by a triple (instead of a single) integral.

Another main drawback of the finite temperature version of the VS scheme concerns the violation of the cusp self-consistency relation, see Eqs.(\ref{FullCusp},\ref{FullCuspVS}), which should be partly responsible for the extended unphysical negative pair correlation function region in the proximity of contact. The importance of simultaneously satisfying the compressibility sum rule and the cusp relation has been manifested by the semi-empirical ESA scheme\,\cite{ESApap1,ESApap2}. Thus, it is worth trying to enforce both self-consistency relations in a dielectric scheme that does not utilize QMC results. With this objective in mind, we bring forth the possibility for an extended VS (eVS) scheme of semi-classical nature that features two self-consistency parameters. The truncation at the first member of the classical BBGKY hierarchy will be based on the eVS closure $f_2(\boldsymbol{r},\boldsymbol{p},\boldsymbol{r}^{\prime},\boldsymbol{p}^{\prime},t)=f(\boldsymbol{r},\boldsymbol{p},t)f(\boldsymbol{r}^{\prime},\boldsymbol{p}^{\prime},t)g(\boldsymbol{r},\boldsymbol{r}^{\prime},t)$ where the non-equilibrium pair correlation function is now given by
\begin{widetext}
\begin{align*}
g(\boldsymbol{r},\boldsymbol{r}^{\prime},t)=g_{\mathrm{eq}}(|\boldsymbol{r}-\boldsymbol{r}^{\prime}|;n,T)+{\alpha}\left[\delta{n}(\boldsymbol{r},t)+\delta{n}(\boldsymbol{r}^{\prime},t)\right]\frac{\partial{g}_{\mathrm{eq}}(|\boldsymbol{r}-\boldsymbol{r}^{\prime}|;n,T)}{\partial{n}}+\gamma\left[\delta{T}(\boldsymbol{r},t)+\delta{T}(\boldsymbol{r}^{\prime},t)\right]\frac{\partial{g}_{\mathrm{eq}}(|\boldsymbol{r}-\boldsymbol{r}^{\prime}|;n,T)}{\partial{T}}\,.
\end{align*}
\end{widetext}
The eVS closure is characterized by the addition of temperature perturbations on an equal footing with the density perturbations. This addition will have profound effects in the mathematical treatment of the linearized kinetic equation and thus in the inner loop of the eVS scheme's algorithm. In particular, the linear temperature response function $\chi_{\mathrm{T}}(\boldsymbol{k},\omega)=\delta{T}(\boldsymbol{k},\omega)/\delta{U}_{\mathrm{ext}}(\boldsymbol{k},\omega)$ would need to be studied in parallel with the linear density response function $\chi(\boldsymbol{k},\omega)=\delta{n}(\boldsymbol{k},\omega)/\delta{U}_{\mathrm{ext}}(\boldsymbol{k},\omega)$, which would require that the zeroth and second moments of the linearized kinetic equation are considered simultaneously, forming a set of equations that allows the determination of both $\chi_{\mathrm{T}}(\boldsymbol{k},\omega)$ and $\chi(\boldsymbol{k},\omega)$. An ideal temperature response, i.e. the second moment equivalent of the ideal (Lindhard) density response, would also naturally emerge. The two self-consistency parameters $\alpha$, $\gamma$ would be determined in the outer loop of the algorithm by a $2\times2$ set of equations that is formed when simultaneously imposing the cusp relation and compressibility sum rule. When $\gamma=0$, it is evident that the envisaged eVS scheme collapses to the standard VS scheme. Finally, it is also worth noting that the coupling between density and temperature fluctuations in the eVS scheme would lead to a density response function expression that does not comply with the general result of the polarization potential approach\,\cite{IchimaB}, see Eq.(\ref{densityresponseDLFC}). On the other hand, since the zero frequency moment sum rule and the quantum fluctuation--dissipation theorem will remain intact, this coupling between density and temperature fluctuations will have no effect on the infinite Matsubara series expression that connects the static structure factor with the density response function, see Eq.(\ref{Matsubaraseries}), which constitutes the backbone of all schemes of the dielectric formalism.

\section*{Acknowledgments}

This work was partially supported by the Center for Advanced Systems Understanding (CASUS) which is financed by Germany's Federal Ministry of Education and Research (BMBF) and the Saxon state government out of the State budget approved by the Saxon State Parliament (CASUS Open Project: \emph{Guiding dielectric theories with ab initio quantum Monte Carlo simulations: from the strongly coupled electron liquid to warm dense matter}). This work has also received funding from the European Research Council (ERC) under the European Union’s Horizon 2022 research and innovation programme (Grant agreement No.101076233, "PREXTREME"). Views and opinions expressed are however those of the authors only and do not necessarily reflect those of the European Union or the European Research Council Executive Agency. Neither the European Union nor the granting authority can be held responsible for them.

\end{document}